\documentclass[aps,prd,twocolumn,amssymb,amsmath,superscriptaddress,floatfix,showpacs,nofootinbib]{revtex4}
\usepackage[dvips]{graphics,graphicx}
\usepackage{multirow}
\usepackage{longtable}
\newcommand{\Whittaker}[3]{G\left(\textstyle #1, #2, #3 \right)}
\begin{document}
\title{Heavy quarkonium correlators at finite temperature:  QCD sum rule approach}
\author{Kenji Morita}
\email{k.morita@gsi.de}
\altaffiliation[Present address: ]{GSI, Helmholtzzentrum f\"{u}r
Schwerionenforschung, Plankstr. 1, 64291 Darmstadt, Germany}
\author{Su Houng Lee}
\email{suhoung@phya.yonsei.ac.kr}
\affiliation{Institute of Physics and Applied Physics, Yonsei
University, Seoul 120-749, Korea}
\date{\today}
\begin{abstract}
 We investigate the properties of heavy quarkonia at finite temperature in
 detail using QCD sum rules. Extending previous analyses, we take into
 account a temperature dependent effective continuum threshold and derive
 constraints on the mass, the width, and the varying effective continuum
 threshold. We find that at least one of these quantities of a
 charmonium changes abruptly in the vicinity of the phase transition.
 We also calculate the ratio of the imaginary time correlator to its
 reconstructed one, $G/G_{\text{rec}}$, by constructing a
 model spectral function and compare it to the corresponding lattice QCD results. We
 demonstrate that the almost constant unity of $G/G_{\text{rec}}$ can be obtained
 from the destructive interplay of the changes in each part of the spectral modification
 which are extracted from QCD sum rules.
\end{abstract}
\pacs{14.40.Gx,11.55.Hx,12.38.Mh,24.85.+p}

\maketitle

\section{Introduction}

In-medium property of heavy quarkonia provides information on the
confinement-deconfinement transition in QCD. In relativistic
heavy ion collisions, final yields measured through dilepton channels depend
on whether they can exist as bound states or not. If the deconfined
plasma is produced, color Debye screening will melt the quarkonia then
suppress the resultant yields~\cite{Matsui_PLB178}. Although $J/\psi$
has been measured in heavy ion collisions at various energies, no
quantitative understanding has been reached yet, because of the
intrinsic complexities of processes in the heavy ion collisions.
See Refs.~\cite{satz06,rapp:_charm,kluberg:_color} for recent reviews.
Therefore, it is important to investigate the properties of the quarkonia
in an ideal environment to give a solid foundation, not
only on the existence of a bound state but also on their detailed spectral
modification such as the mass shift and broadening at finite temperature or density.
In this respect, it was pointed out that a downward mass shift of $J/\psi$ in
hadronic matter was caused by the decrease of string tension, which had
been predicted by lattice QCD, and as such can be a precursor phenomena of the
confinement-deconfinement transition \cite{hashimoto86}. Therefore,
the detailed determination of the spectral properties can play a key role
in the study of QCD phase transition.

Properties of the bound states have been traditionally investigated with
quantum mechanical potential models. It is known that the mass spectrum
of heavy quarkonium can be described well by the so-called Cornell potential, which
implements the Coulomb potential at short distance and linearly rising
one at long distance \cite{eichten78:_charm,eichten80}.
This approach can be extended to finite temperature by assuming
that all the effects of temperature can be accounted for by the temperature dependent potential \cite{mocsy:_poten}. In this approach, however, how to
construct the potential relevant for Schr\"{o}dinger equation is a
non-trivial problem and various types have been examined by
incorporating properties known from lattice QCD \cite{Wong05,alberico05,mocsy06,cabrera07:_t,mocsy08}.
Although the potential model approach is related to QCD only through
the temperature-dependent potential computed with lattice QCD, recent
development in weak coupling methods such as pNRQCD
\cite{brambilla00:_poten_nrqcd,brambilla05,brambilla08} which is an effective
field theory of QCD and a resummed perturbative approach
\cite{laine07,laine07:_qcd} shed a light on more rigorous foundation of
the heavy quark potential at finite temperature.

Direct evaluation of the quarkonium properties with lattice QCD has been
carried out through the maximum entropy method (MEM)
\cite{Asakawa_PRL92,Datta_PRD69,umeda05}. One can reconstruct the spectral
function of a given channel by inverting the dispersion relation of the current
correlation function calculated in the imaginary time. Indeed,
Ref.~\cite{Asakawa_PRL92} indicated the existence of
$J/\psi$ bound state even in the deconfined phase up to $T\sim 1.6T_c$.
At high temperature, however, lattice QCD suffers from the limited size of
the temporal direction, which also affects the accuracy of the reconstruction of
the spectral function in MEM \cite{umeda05,jakovac07}. In this approach, the only reliable information seems to be the presence or disappearance of the first peak in the spectral density. A potential model calculation
\cite{mocsy07,mocsy08} indicates that the first peak observed in MEM at
high temperature can be attributed to the threshold enhancement and that
$J/\psi$ has already melted at $T=1.2T_c$.

Recently AdS/QCD approach has also been applied to the heavy quarkonium in
medium \cite{kim07:_heavy_qcd,fujita:_finit_ads_qcd}. Although a direct relation to real QCD is still missing, the approach seems
to give another insight to the problem from the viewpoint of the
strongly coupled gauge field theory.
Both Refs.~\cite{kim07:_heavy_qcd,fujita:_finit_ads_qcd} show notable
spectral change in $J/\psi$ around and above $T_c$.

In previous works
\cite{morita_jpsiprl,morita_jpsifull,song09,lee_morita_song_prep}, we
have proposed another approach to study the properties of heavy
quarkonia at finite temperature based on QCD sum rules; the approach extended the
previous studies at nuclear medium~\cite{Klingl_PRL82,Hayashigaki99:_jpsi}.
The QCD sum rule \cite{Shifman_NPB147,Shifman_NPB147_2} provides a
systematic framework which connects
the current correlation function at deep Euclidean region
to the spectral function integrated with respect to the energy variable
with a weight that makes the integral dominated by the lowest pole. It has
been applied to various aspects of
hadrons quite successfully \cite{reinders85,Narison_sumruletextbook}.
Due to the asymptotic freedom, one can reliably compute the correlation
function at the deep Euclidean region using perturbation theory
via the operator product expansion (OPE) which provides non-perturbative
correction through QCD condensates. For a heavy quarkonium, to a good
approximation, one can truncate the expansion at the lowest dimensional
local operator, which is the dimension four gluon condensate.

The aim of this work is to extend our previous works to a more systematic
analysis by incorporating the continuum part of the model spectral function,
applying a more sophisticated optimization procedure in determining the
spectral parameters, and then
making a  comparison to the lattice QCD results. In this paper, we
describe the detailed procedure based on the Borel transformation which is
widely used in QCD sum rule applications.
Then we discuss the spectral change of charmonia and bottomonia at
finite temperature near and above $T_c$. Using the spectral parameters
obtained in the QCD sum rules, we construct model spectral functions
and compute the imaginary time correlators, which we will compare with
lattice QCD.

This paper is organized as follows. In the next section, we briefly
review the QCD sum rules for heavy quarkonium at finite temperature,
then explain the procedure based on the Borel transformation.
We will show the results of the spectral parameters in
Sec.~\ref{sec:spectral_result}. We will discuss the imaginary time
correlators reconstructed from the spectral parameters in Sec.~\ref{sec:IMC}.
Section \ref{sec:summary} is devoted to the summary.

\section{QCD sum rules for heavy quarkonium in medium}

\subsection{OPE for correlation function}

We start with the current correlation function
\begin{equation}
 \Pi^J(q) = i\int d^4 x e^{iq\cdot x} \langle T[J^J(x)J^J(0)] \rangle.\label{eq:correlation}
\end{equation}
We choose the currents for pseudoscalar ($P$), vector ($V$), scalar
($S$), and axial-vector ($A$) as
\begin{align}
 j^P&= i\bar{h}\gamma_5 h\\
 j^V_\mu& = \bar{h}\gamma_\mu h\\
 j^S&= \bar{h}{h} \\
 j^A_\mu&= \left(q_\mu q_\nu / q^2-g_{\mu\nu}\right)\bar{h}\gamma_5
 \gamma^\nu h
\end{align}
with $h$ being the heavy quark operator, $c$ or $b$. For the axial-vector, we
pick up the conserved components for $\chi_{c(b)1}$ states.
We will take the expectation value at finite temperature.
Therefore, in general, there are two independent components in both the $V$ and
the $A$ channels.
We assume the momentum of the current to be $q^\mu = (\omega,\boldsymbol{0})$,
\text{i.e.}, a pair of quark and antiquark at rest with respect to the medium such
that only one component becomes independent. Then we define the
dimensionless correlation function as
\begin{align}
 \tilde{\Pi}^{P,S}(q^2)&= \frac{\Pi^{P,S}(q)}{q^2}\\
 \tilde{\Pi}^{V,A}(q^2)&= \frac{\Pi^{\mu, V,A}_\mu(q)}{-3q^2}
\end{align}
which can be expanded up to dimension four operators via OPE
\begin{equation}
 \tilde{\Pi}^J(q^2) \simeq C^J_1 + C^J_{G_0}G_0 + C^J_{G_2}G_2.\label{eq:ope}
\end{equation}
with $G_0=\langle \frac{\alpha_s}{\pi}G^a_{\mu\nu}G^{a,\mu\nu} \rangle$ and $G_2$
being the scalar and twist-2 gluon condensates, respectively.
$G_2$ is defined as the traceless and symmetric part of the gluon
operator as
\begin{equation}
 \left\langle\frac{\alpha_s}{\pi}G^a_{\mu \alpha}
  G_\nu^{a,\alpha}\right\rangle = \left(u_\mu u_\nu - \frac{1}{4}g_{\mu\nu}\right)G_2
\end{equation}
which vanishes at $T=0$ according to the Lorentz invariance.
The medium four velocity $u^\mu$ is set to $(1,0,0,0)$.
Hereafter we assume that all medium effects are imposed on the change of
the local operators \cite{Hatsuda93}. This assumption is justified when
the typical scale of the condensates is smaller than the separation scale
\cite{lee_morita_song_prep}, namely,
\begin{equation}
 4m_h^2-q^2 \gg \langle G \rangle \sim (\Lambda_{\text{QCD}}+aT+b\mu)^2.\label{eq:sep_scale}
\end{equation}
with $m_h=m_c$ or $m_b$ being the heavy quark mass. The large heavy quark mass $m_h$ allows us to work even at the physical energy scale
$q^2 = m^2_{J/\psi}, m_{\Upsilon}^2$ and so on, although somewhat
marginal in reality \cite{peskin79}. In this case, the OPE
gives a formula for the bound state mass that is proportional to  the change of the color electric field squared.  This formula is the QCD second order Stark effect
\cite{peskin79,luke92,lee_morita_stark} that can also be obtained from the
leading order correction of the static potential in pNRQCD~\cite{brambilla08}.
Combining the formula with the temperature dependence of the electric
condensate, one finds a sudden mass shift at
$T_c$~\cite{lee_morita_stark}.
In the QCD sum rule, we go to
the deep Euclidean region $q^2 = \omega^2 = -Q^2 \ll 0$, in which
the condition \eqref{eq:sep_scale} is well fulfilled.
Therefore the Wilson coefficients $C_i$ are the same as the vacuum case
and have been calculated in
Refs.~\cite{Reinders_NPB186,reinders85,Klingl_PRL82,song09}. See
Ref.~\cite{lee_morita_song_prep} for a list.

\subsection{Borel transformation and dispersion relation}

The Borel transformation of the correlation function is defined as
\begin{equation}
 \mathcal{M}^J(M^2)  =\lim_{\substack{Q^2/n \rightarrow M^2, \\ n,Q^2
  \rightarrow \infty}}
  \frac{(Q^2)^{n+1}\pi}{n!}\left(-\frac{d}{dQ^2}\right)^n
  \tilde{\Pi}^J(Q^2).
\end{equation}
If one does not take the limit, the derivative of the correlation function
corresponds to the moment of the correlation function which was used
in the previous sum rule works for the heavy quarkonia
\cite{Shifman_NPB147_2,Reinders_NPB186,Klingl_PRL82,Hayashigaki99:_jpsi,morita_jpsiprl,morita_jpsifull,song09}.
Taking this limit corresponds to going to deeper Euclidean region for
better perturbative expansion while retaining the connection to the resonance
through large $n$ \cite{Shifman_NPB147}. Indeed,
Eq.~\eqref{eq:sep_scale} is expected to be better satisfied as the typical OPE term $\frac{1}{(4m_h-q^2)^d} \langle G^d \rangle$  reduces to
$\frac{1}{d!M^{2d}}e^{-4m_h^2/M^2}  \langle G^d \rangle$ after the transformation,
and hence the condensate contribution is further suppressed by $\frac{1}{d!}$.
In the heavy quarkonia, the moment
sum rule works well enough to extract the mass due to the large
separation scale coming from the heavy quark mass. Nevertheless, the
Borel transformation approach has several advantages for more systematic
analysis as revealed below.

\begin{figure}[!t]
 \includegraphics[width=3.375in]{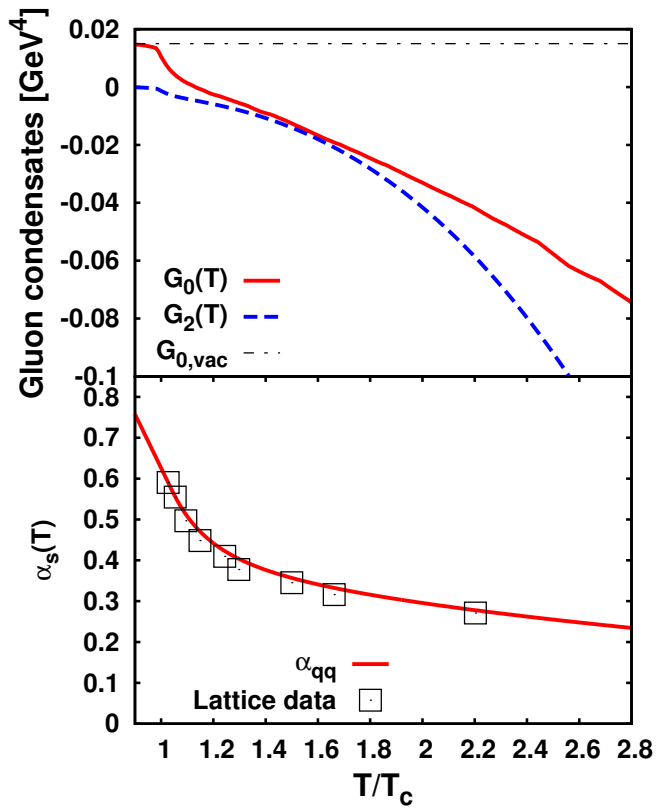}
 \caption{(Color online). Temperature dependence of the gluon
 condensates $G_0(T)$ and $G_2(T)$ (top) and effective temperature
 dependent coupling constant (bottom).}
 \label{fig:gc}
\end{figure}

For the expanded heavy quarkonium correlation function \eqref{eq:ope},
the Borel transformation can be analytically carried out as
\begin{align}
 \mathcal{M}^J(M^2) &= e^{-\nu}\pi
  A^J(\nu)[1+\alpha_s(M^2)a^J(\nu)+b^J(\nu)\phi_b(T) \nonumber\\
 &+c^J(\nu)\phi_c(T)]
 \label{eq:borel_moment}
\end{align}
with a dimensionless scale parameter $\nu=4m_h^2/M^2$.
The first line of Eq.~\eqref{eq:borel_moment} is the same as that
derived in Ref.~\cite{bertlmann82} except for the temperature dependency
of the scalar gluon condensate term $\phi_b$. The second line shows the
twist-2 term which appears in the case of medium. $\phi_b$ and $\phi_c$ are
defined as
\begin{align}
 \phi_b&= \frac{4\pi^2}{9(4m_h^2)^2}G_0(T),\label{eq:phib}\\
 \phi_c &= \frac{4\pi^2}{3(4m_h^2)^2}G_2(T),\label{eq:phic}
\end{align}
as given in Ref.~\cite{morita_jpsiprl,morita_jpsifull}. While
$A^J(\nu)$, $a^J(\nu)$ and $b^J(\nu)$ are given in Ref.~\cite{bertlmann82},
the transformed twist-2 coefficient $c^J(\nu)$ is derived for the first
time in this paper.
For completeness, we list all the
Borel transformed Wilson coefficients used in
Eq.~\eqref{eq:borel_moment} in Appendix \ref{app:wilson}. While
Bertlmann worked on the
on-shell renormalization of heavy quark mass in Ref.~\cite{bertlmann82},
we maintain the off-shell
renormalization as a straightforward extension from
Ref.~\cite{Reinders_NPB186}. Hence, the correction term
$ -\frac{4\ln 2}{\pi}h^J(\nu) $ in $a^J(\nu)$ is included throughout
this calculation. Note that this term is also necessary to keep the
perturbative correction term small enough in $\mathcal{M}(M^2)$ and in
$-\frac{\partial \mathcal{M}(M^2)}{\partial(1/M^2)}$ that is used later.
Since this part is temperature independent, the
difference does not affect our aim but enables us to proceed in a more
transparent way by retaining the relation with the previous moment
sum rule analyses.
In Eq.~\eqref{eq:borel_moment}, external inputs are
heavy quark mass $m_h$ (contained in $A(\nu)$, $\phi_b$ and $\phi_c$),
strong coupling constant $\alpha_s(M^2)$ and gluon condensates $G_0(T)$ and $G_2(T)$.
In this paper, we put $m_c(p^2=-m_c^2)=1.262$ GeV inferred from the
$m_c(p^2=-2m_c^2)=1.24$ GeV used in the previous works
\cite{Klingl_PRL82,morita_jpsiprl,morita_jpsifull,song09,lee_morita_song_prep}
and $m_b(p^2=-m_b^2)=4.12$ GeV. $\alpha_s(M^2)$ is calculated from the
running coupling formula from $\alpha_s(8m_c^2)=0.21$ also used in
the previous works. $\alpha_s(8m_b^2)=0.158$ is used for bottomonia.
The gluon condensates have been extracted from the
results of pure SU(3) lattice gauge theory \cite{Boyd_NPB469} and shown
in Fig.~\ref{fig:gc}. Here effective temperature dependent coupling
constant $\alpha_s(T)$ is used for the determination of $G_2(T)$ which
is the symmetric and traceless part of the gluon operator.
This is done with the identification $\langle
\frac{\alpha_s}{\pi}\mathcal{ST}(G^a_{\alpha
\mu}G^{a,\mu}_\beta)\rangle \equiv \frac{\alpha_s(T)}{\pi}\langle
\mathcal{ST}(G^a_{\alpha\mu}G^{a,\mu}_\beta) \rangle$ based on the
the separation scale in the OPE.
We adopted $\alpha_{qq}(T)$ shown Ref.~\cite{Kaczmarek_PRD70} and took
the value at $r=r_{\text{screen}}$. It is also shown in the bottom
panel of Fig.~\ref{fig:gc}.

The current correlation function is related to the spectral function
through the dispersion relation. At finite temperature, the spectral
function is given by the imaginary part of the retarded correlation function
$\Pi^R(q)$ which is in general different from
Eq.~\eqref{eq:correlation}. However, we can relate it to the spectral
function by virtue of the fact
that $\Pi^{R}(\omega)=\Pi(\omega^2)$ since $\omega^2 = -Q^2 < 0$
\cite{Hatsuda93}. Putting $\tanh(\sqrt{s}/2T)=1$, which is safely
satisfied for $\sqrt{s}\sim m_{h\bar{h}}$, we have the dispersion relation for the
Borel sum rule in the same form as the vacuum case
\begin{equation}
 \mathcal{M}^J(M^2) =
  \int_{0}^{\infty}ds\,e^{-s/M^2}\text{Im}\tilde{\Pi}^J(s). \label{eq:dispersion}
\end{equation}
Note that the weight factor in the dispersion integration is now
exponential $e^{-s/M^2}$ while it was the inverse power $(s+Q^2)^{-n}$ in the moment sum rule.

\subsection{Analysis procedure}
\label{subsec:method}

Assuming the quark-hadron duality, we take a model spectral function of
a given current as a simple ansatz for the imaginary part of the
correlation function and call it the phenomenological side. First
we decompose it into the pole and the continuum contribution as
\begin{equation}
 \text{Im}\tilde{\Pi}(s) = \text{Im}\tilde{\Pi}^{\text{pole}}(s) + \text{Im}\tilde{\Pi}^{\text{cont}}(s),\label{eq:phen}
\end{equation}
with
\begin{align}
 \text{Im}\tilde{\Pi}^{\text{pole}}(s)&= \begin{cases}
				       \displaystyle f_0\delta(s-m^2),
				       \quad \Gamma=0\label{eq:phen_pole}\\
				       \displaystyle \frac{f\Gamma
				       \sqrt{s}}{(s-m^2)^2+s\Gamma^2},\quad
				       \Gamma > 0, \quad s > 4m_h^2
				      \end{cases},\\
 \text{Im}\tilde{\Pi}^{\text{cont}}(s)&=
 \theta(s-s_0)\text{Im}\tilde{\Pi}^{J,\text{pert}}(s).
 \label{eq:phen_cont}
\end{align}
The pole term is the same as in the previous works
\cite{morita_jpsiprl,morita_jpsifull,song09}. We consider possible finite
width in the deconfined medium, because decay into $h\bar{h}$ pair,
which was forbidden in vacuum due to the Okubo-Zweig-Iizuka (OZI) rule,
becomes possible. This is done by implementing a relativistic
Breit-Wigner function and cutting off the
contribution below $h\bar{h}$ threshold to avoid possible numerical artifacts as
discussed later in Sec.~\ref{subsec:charm}.

We adopt the perturbative
part of the spectral function including $\alpha_s$ correction but with
the sharp threshold factor $\theta(s-s_0)$ as a
model for the continuum; such form reproduces the corresponding part of the
OPE side when putting $s_0=4m_h^2$. These functional forms are explicitly given in
Ref.~\cite{reinders85} and listed in Appendix \ref{app:cont} for the completeness. Since there are known excited states such as
$\psi'$ between the lowest lying state and the physical continuum threshold, one
may think this model is an oversimplification of the real
spectrum. However, due to the suppression coming from
the Borel transformation, this simplification does not affect
the property of the lowest pole. Instead, this form results in a little
smaller continuum threshold value than that from the analysis incorporating the excited
states explicitly as we will see later.
Moving the continuum part to the OPE side in Eq.~\eqref{eq:dispersion},
one can isolate the pole term. There is an additional contribution to the
spectral function from the absorption of the current by the thermally
excited particle, \textit{i.e.}, Landau damping which shows up as a peak
at $s=0$ when $\boldsymbol{q}=0$. This was recognized in
Ref.~\cite{Bochekarev86} in a QCD sum
rule framework and has been called the ``scattering term''. Recently it has been
emphasized that this gives constant contribution to the imaginary time
correlator which is the basis of the spectral function study in the
lattice QCD \cite{umeda07}. In the QCD sum rule application in the
deconfined phase, we can neglect this contribution, as
explained in Ref.~\cite{song09,lee_morita_song_prep}. The scattering
term appears in the OPE through the bare heavy quark condensates
$\langle \bar{h} \Gamma DD..D h \rangle$ which is converted into the gluon condensates
via heavy quark expansion at $T=0$ \cite{generalis84}. To a first
approximation that assumes free heavy quarks in a medium, we can put
the same quantity on the phenomenological side so that it cancels the corresponding contribution in the OPE.

Differentiating both side of
Eq.~\eqref{eq:dispersion} with respect to $1/M^2$ and taking its ratio
to the original equation, one has
 \begin{gather}
  -\frac{\displaystyle
   \frac{\partial}{\partial(1/M^2)}
   [\mathcal{M}(M^2)-\mathcal{M}^{\text{cont}}(M^2)]}
   {\mathcal{M}(M^2)-\mathcal{M}^{\text{cont}}(M^2)} \nonumber\\
   = \frac{ \displaystyle \int_{4m_h^2}^{\infty} ds \, s\,e^{-s/M^2}
   \text{Im}\tilde{\Pi}^{\text{pole}}(s)}
   {\displaystyle \int_{4m_h^2}^{\infty} ds \, e^{-s/M^2}
   \text{Im}\tilde{\Pi}^{\text{pole}}(s)}\label{eq:sumrule}
 \end{gather}
where $\mathcal{M}^{\text{cont}}(M^2)$ is the Borel-transformed
continuum spectral function according to Eq.~\eqref{eq:dispersion}. One
immediately finds that the right-hand side of Eq.~\eqref{eq:sumrule}
gives the squared pole mass $m^2$ for $\Gamma=0$. There are three
spectral parameters to be determined in Eq.~\eqref{eq:sumrule}:
pole mass $m$, width $\Gamma$ and effective continuum threshold $s_0$.
The strength parameter $f_0$ or $f$ contained in the pole term cancels
by taking the ratio. We solve Eq.~\eqref{eq:sumrule} for the mass $m$ as a
function of the Borel mass $M^2$ (which we call Borel curve) with given
sets of $\Gamma$ and $s_0$. While the extracted mass $m$ depends on
$M^2$ through the left-hand side of Eq.~\eqref{eq:sumrule} by
construction, it should not do so since $M^2$ is an external
parameter. The apparent dependence of $m$ on $M^2$ is due to the truncation
of the OPE and the insufficient subtraction of excited states and
the continuum part in the spectral density. The truncation of the OPE
shows up as strong $M^2$ dependency at small $M^2$ while the
insufficient subtraction does at large $M^2$. In
practice, however, we can expect $M^2$ independent $m$ at intermediate
$M^2$ region after tuning the effective threshold parameter $s_0$ in the continuum
part \cite{hatsuda95:_qcd,Hatsuda93}. Introducing the finite
width also affects the $m(M^2)$
in the small $M^2$ region \cite{lee_morita_nielsen_exoticwidth}.
We thus introduce the following quantity to determine the best set
of the parameters that gives the flattest curve in the intermediate
$M^2$ region;
\begin{equation}
 \chi^2 \equiv
  \frac{1}{M^2_{\text{max}}-M^2_{\text{min}}}\int_{M^2_{\text{min}}}^{M^2_{\text{max}}}dM^2 [m(M^2)-m(M^2_0)]^2\label{eq:chi2}
\end{equation}
where $M_0^2$ is defined by $dm(M^2)/dM^2|_{M^2=M^2_0}=0$.
The range of the intermediate $M^2 \in [M_{\text{min}}^2,
M_{\text{max}}^2]$ is called Borel window,
in which the convergence of the OPE and the pole dominance of the
dispersion integral are satisfied.
We fix $M_{\text{min}}^2$ by requiring
the dimension four correction terms to be smaller than 30\% of
the total since it is expected that the contribution from the next higher dimensional
operator is kept less than 10\% of the total within this condition
\cite{Shifman_NPB147}. For bottomonium systems, while this condition
is always fullfilled in typical $M^2$ values due to the larger quark
mass in Eqs.~\eqref{eq:phib} and \eqref{eq:phic}, the perturbative
radiative correction term $\alpha_s(M^2)a^J(\nu)$ can be large enough to
spoil the perturbative expansion. We therefore impose this term to be
less 0.3 at $M^2 > M_{\text{min}}^2$.
The pole dominance should be also imposed
on the criteria to preserve the reliability on the extracted property of the pole part
of the spectral density. We determine $M^2_{\text{max}}$ by requiring the continuum
contribution to be less than 30\% of the total perturbative term. As we
shall see, reducing the value of the continuum threshold makes the Borel
curve flatter. Therefore, this criterion does not affect the pole mass
at the best
fit while it becomes important when the Borel window is narrow.
The $\chi^2$ measures the average deviation of
$m(M^2)$ from its value at the $M^2$ ``plateau'' characterized by
$dm(M^2)/dM^2|_{M^2=M^2_0}=0$. While it vanishes in the case of
completely $M^2$ independent mass, $\sqrt{\chi^2}$ can be regarded as
a systematic error on the extracted mass $m(M_0^2)$ such that
$m(M_0^2)+\sqrt{\chi^2}$ gives the upper limit. Note, however, that it
gives much larger deviation than actual uncertainty when Borel
window contains the strongly $M^2$ dependent part of the Borel
curve at small $M^2$ which is due to the truncation of OPE. Hence, a refinement
on the determination of the Borel window will give more quantitative
insights on the uncertainty in the extraction of the mass parameter.

\begin{figure}[t]
 \includegraphics[width=3.375in]{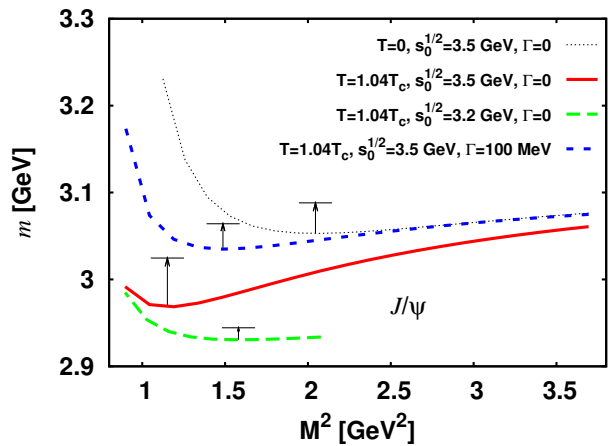}
 \caption{(color online). Some examples of the Borel curve taken from $J/\psi$ at
 $T=1.04T_c$. Solid line shows the curve for $\sqrt{s_0}=3.5$ GeV and
 $\Gamma=0$. Long-dashed line shows the case in which $\sqrt{s_0}$ is reduced
 while in short-dashed line $\Gamma$ is increased to 100 MeV. For
 reference, the thin dotted line shows the curve at $T=0$ with
 $\sqrt{s_0}=3.5$ GeV and $\Gamma=0$. The arrows accompanied by a short
 horizontal line indicates the upper limit of the mass at $M^2=M_0^2$
 evaluated by $\sqrt{\chi^2}$.}
 \label{fig:borelexample}
\end{figure}

\begin{figure}[tb]
 \includegraphics[width=3.375in]{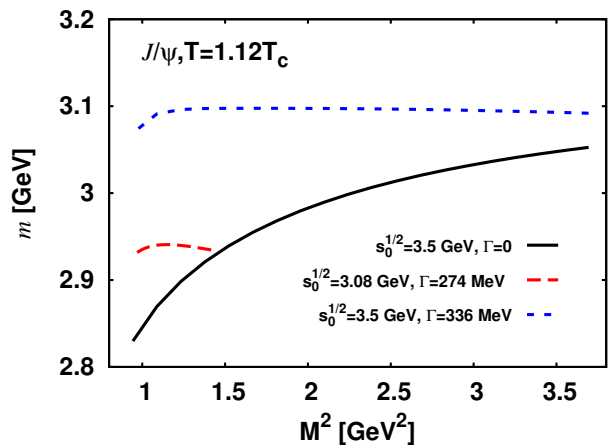}
 \caption{(color online). Same as Fig.~\ref{fig:borelexample}, but for
 $T=1.12T_c$. See text for description.}
 \label{fig:borelexample2}
\end{figure}

To illustrate the minimization procedure, we show below how the Borel
curve obtained by solving Eq.~\eqref{eq:sumrule} changes with
respect to the external parameters. We depict some examples in
Fig.~\ref{fig:borelexample}.
Let us start with the dotted curve corresponding to $T=0$,
$\sqrt{s_0}=3.5$ GeV, and $\Gamma=0$ for $J/\psi$.
>From the definition of the Borel mass
$M^2$, one sees that the Borel curve looks like the $n$ dependence of
the moment ratio shown in Refs.~\cite{morita_jpsiprl,morita_jpsifull}
except for the reversed direction of
the horizontal axis. Here we draw the lines only within the
Borel window. Therefore the dotted line is truncated at $M^2= 1.12$ GeV$^2$ and $3.69$
GeV$^2$.  The arrow on each line indicates the location of $M_0^2$ and
the upper limit evaluated from $\sqrt{\chi^2}$.
As temperature increases, the gluon condensates decrease as shown
in Fig.~\ref{fig:gc}. This is reflected by the
lowered curves obtained for $T=1.04T_c$, as in line with the moment sum rule case
\cite{morita_jpsiprl,morita_jpsifull}. Since the condensate contribution becomes
dominant in the OPE side at small $M^2$, one sees great reduction of mass in this
region. Then the solid curve, for $T=1.04T_c$, shows a minimum at
smaller Borel mass. This corresponds to the shift of the minimum of the
moment ratio to large $n$ in the moment sum rule.
Regarding the solid curve as the base line, one sees that decreasing the
continuum threshold flattens the curve at large $M^2$. In this case,
however, reduction of the continuum threshold makes the
$M^2_{\text{max}}$ smaller and thus the resultant Borel window becomes
narrower. This is why the dash line ends at $M^2_{\text{max}}=2.08$
GeV$^2$. One also sees that the curve is almost
flat above $M^2 = 1.5$ GeV$^2$. This means that further reduction of the
continuum threshold breaks the stability of the Borel curve, \textit{i.e.},
the mass decreases monotonically and no minimum would exist.
$M^2_{\text{min}}$ does not change against the reduction of the continuum threshold
since it depends only on temperature through the power correction terms
in the OPE. On the other hand, if one increases the width, it raises
the mass, especially at low $M^2$ as clearly seen in the short-dashed
curve. One sees that, at $M^2$ far from the $M^2_{\text{min}}$, the
two lines, one obtained by reducing the continuum threshold
and the other by introducing the width, show almost similar flatness. In
the present case,
the rapid rise in the $\Gamma=100$ MeV gives $\chi^2=8.55\times10^{-4}$ GeV$^2$
which is much bigger than $\chi^2=1.89\times10^{-4}$ GeV$^2$ of the
$\sqrt{s_0}=3.2$ GeV curve. However, this depends on the choice of the
criterion in the determination of the Borel window. If one tightens the
criterion, to 10\% power correction for instance, $M^2_{\text{min}}$
becomes larger and then $\chi^2$ of the $\Gamma=100$ MeV will be
smaller.
This indicates the difficulty in accurately  determining the spectral
parameters when one takes into account the change of both the width and the
continuum threshold. Nevertheless, one
can make the curve flatter by decreasing the continuum threshold without
introducing broadening up to a certain temperature.
Note that the arrows become shorter as the curve
becomes flatter. As explained above, while there is about 50 MeV
uncertainty in the largest case, it is due to the strong $M^2$
dependence seen in small $M^2$ region. Therefore one should not take
these values so seriously.

This situation changes if one goes to higher temperatures. We plot some
examples from $T=1.12T_c$ in Fig.~\ref{fig:borelexample2}. As the solid
line shows, no stability is achieved in $\sqrt{s_0}=3.5$ GeV and
$\Gamma=0$ case. This is similar to what is seen in the moment sum rule above
$T > 1.05T_c$ \cite{morita_jpsiprl}. Now we can try to restore the
stability by varying $s_0$ and $\Gamma$. From what we learned from
Fig.~\ref{fig:borelexample}, reducing $s_0$ decrease $m(M^2)$ especially
at high $M^2$. In this case, however, Borel window closes before
stability is restored; $M^2_{\text{max}} < M^2_{\text{min}}$ occurs.
Therefore, one has to increase width to recover the stability.
In other words, the breakdown of the stability occurring above $T_c$ can
now be regarded as the onset of the broadening. We denote this
temperature as $T_{\text{onset}}$, which depends on the channel as we
shall see below. Note that this does not mean $\Gamma$ must be 0 below
$T_{\text{onset}}$, since one can find
the best parameter set with $\Gamma > 0$ at $T < T_{\text{onset}}$
after an additional constraint is given. $T_{\text{onset}}$ should be regarded
as the upper limit of temperature at which broadening sets in.
One should also note that $T_{\text{onset}}$ depends on the criteria for the
Borel window. One can broaden the Borel window by relaxing either or
both of the criteria. For example, if one sets the continuum contribution
to be less than 50\% instead of 30\%, $M^2_{\text{max}}$ becomes larger
thus can open the Borel window. Indeed such a
situation is realized in some cases considered in this paper, as seen in
the resultant spectral parameters summarized in
Tables~\ref{tbl:sumruleresult_jpsi}-\ref{tbl:sumruleresult_chib1}.
When the $\chi^2$ takes its minimum at the smallest $\sqrt{s_0}$
satisfying the criteria, making the Borel window
larger by relaxing the $M^2_{\text{max}}$ criterion leads to
smaller $\sqrt{s_0}$ while retaining the width.
Note, however, that there might be ``pseudopeak'' artifact in the Borel
curve for a too relaxed criterion \cite{kojo08:_sigma_qcd}.

The long-dashed line in Fig.~\ref{fig:borelexample2} denotes the case in
which we introduce $\Gamma=274$ MeV with decreasing the continuum
threshold to $\sqrt{s_0}=3.08$ GeV. As a result of raising the Borel
curve at small $M^2$ while lowering it at large $M^2$, the shape of the
curve becomes convex contrary to the lower temperature cases. As seen in the
short-dashed line, one can restore the stability only if one increases the
width. If we keep $\sqrt{s_0}=3.5$ GeV, the resultant width is 336 MeV.
The values of $\chi^2$ are $1.69\times 10^{-5}$ GeV$^{2}$ for $\sqrt{s_0}=3.08$ GeV
and $\Gamma=274$ MeV and $1.9\times10^{-5}$ GeV$^{-2}$ for for $\sqrt{s_0}=3.5$ GeV and
$\Gamma=336$ MeV, respectively, indicating the almost equally flat curves
and again the difficulty of comparing the curves by varying both $s_0$ and
$\Gamma$. In Fig.~\ref{fig:borelexample2}, $\sqrt{\chi^2}$ for the two stable
Borel curves are a few MeV, which are small enough to be neglected in
the figure.

In the following, we use the $\chi^2$ evaluation using
Eq.~\eqref{eq:chi2} only for determining the best curve among those cases with
the same continuum threshold but with different $\Gamma$, in
order to avoid the biases imposed by the choice of the Borel window. In
this way one fixes one edge of the curve $M^2_\text{max}$ so that $\chi^2$
measures only the effect of introducing the width. For some cases where
$\Gamma=0$ always gives the flattest curve, we may use $\chi^2$ to
determine the best $s_0$ value, since the other edge
of the Borel curve is fixed. For example, we can safely determine the best
$s_0$ by evaluating $\chi^2$ at $T=0$. Furthermore, though we maintain the criterion
for the Borel window as explained, we can easily estimate how the best
value changes with respect to the change of the criterion. If one
relaxes the pole dominance condition, it extends the Borel window to
larger $M^2$ therefore continuum threshold giving the best $\chi^2$
will become smaller. On the other hand, if one requires the smaller power
correction, $M^2_{\text{min}}$ becomes larger and thus the $\chi^2$ will be
more sensitive to the continuum. As long as we preserve reasonable values
of these criterion, typically $10-30\%$ for power correction and less
than $50\%$ for the continuum contribution, we find the uncertainty of the
obtained mass to be about a few tens MeV. Since we maintain the same
criterion even at different temperatures, the relative in-medium change of the spectral
parameters is not affected by the particular choice of the criterion for the  Borel window.

\begin{figure}[!t]
 \includegraphics[width=3.375in]{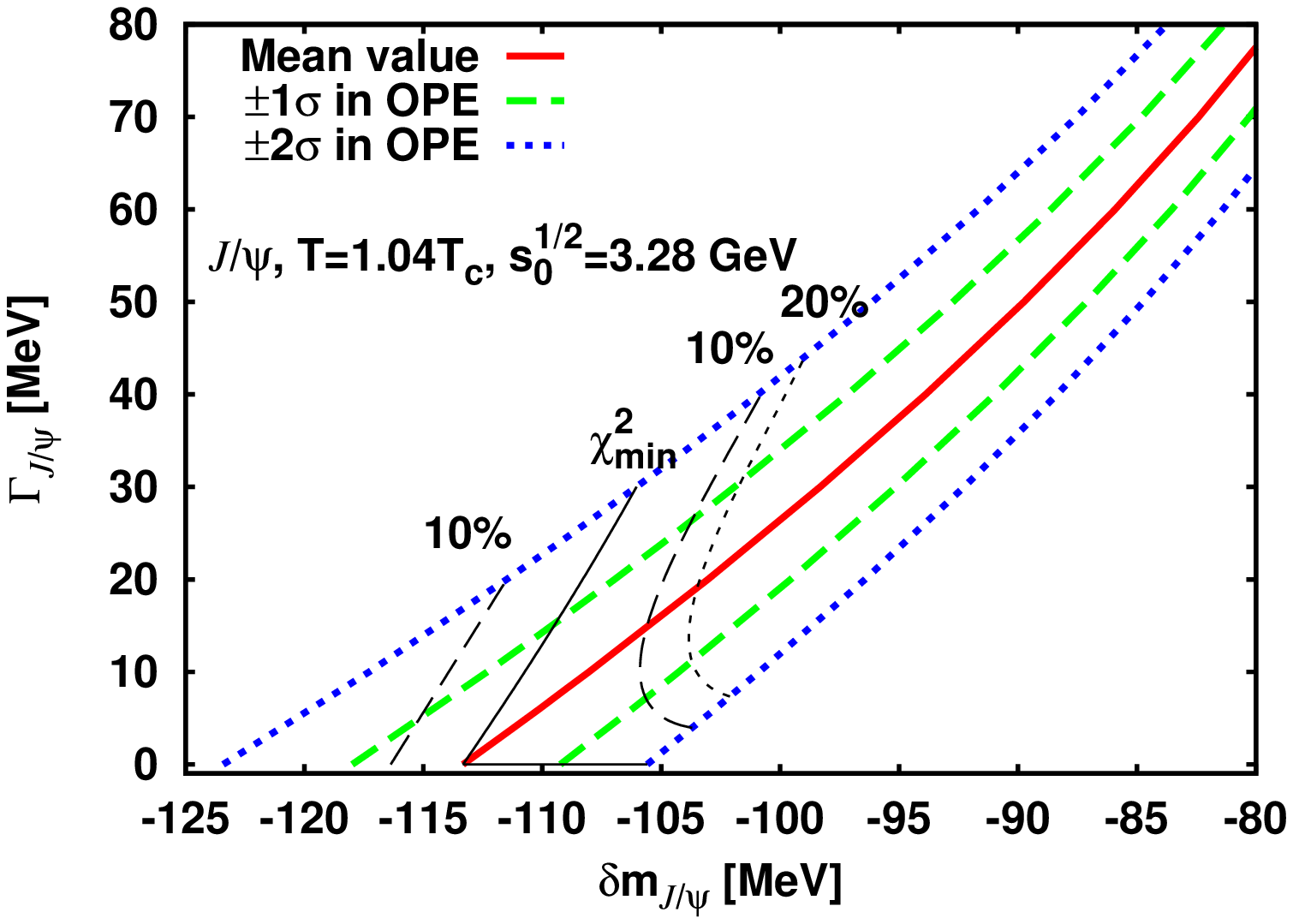}
 \caption{(color online). Constraint on the mass shift and the width of
 $J/\psi$ at $T=1.04T_c$ and $\sqrt{s_0}=3.28$ GeV. In addition to the
 thick lines denoting the constraint, contours of the equal $\chi^2$
 values (unit of deviation from the minimum) are plotted as thin lines.}
 \label{fig:g-m_s328}

 \includegraphics[width=3.375in]{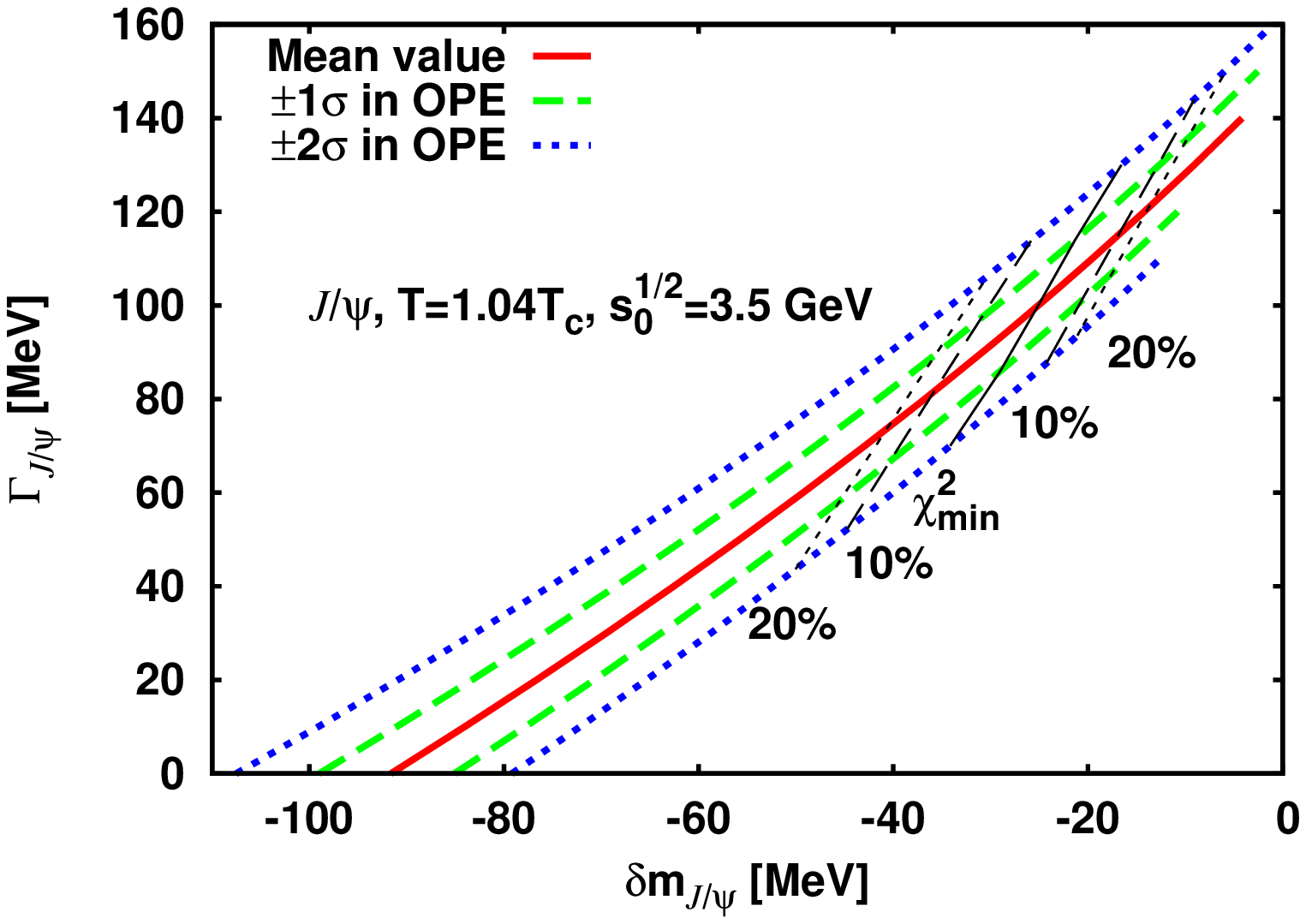}
 \caption{(color online). Same as Fig.~\ref{fig:g-m_s328} but for
 $\sqrt{s_0}=3.5$ GeV. }
 \label{fig:g-m_s350}
\end{figure}

Before closing the section, we would like to address possible
uncertainty on the extracted parameters. The source of the uncertainty
is roughly classified into two parts; one is temperature independent and
the other is dependent quantities. The former consists of the gluon condensate at
$T=0$, strong coupling constant, and the heavy quark masses. Whereas these quantities
certainly affect the value of the spectral parameters, the relative
changes at finite temperatures from the vacuum values do not differ by
changing them within the constraints from the experiment. Hence we focus
on the effect of the temperature dependent part here. We estimate the
$1\sigma$ and $2\sigma$ uncertainty of the OPE side through the
temperature dependent part of the gluon condensates by reading off that
of the energy density and pressure in the lattice results shown in
Ref.~\cite{Boyd_NPB469}. Then, we extract the spectral parameters with
the OPE side shifted by $\pm1\sigma$ and $\pm 2\sigma$. The resultant
constraints which show up the typical correlation between the mass and
the width together with the uncertainty deduced from the deviation of
the OPE side are displayed in Figs.~\ref{fig:g-m_s328} and
\ref{fig:g-m_s350}. The solid line in Fig.~\ref{fig:g-m_s328} shows the case corresponding
to the long-dashed line in Fig.~\ref{fig:borelexample}; the best Borel
curve is obtained by reducing the threshold parameter without
introducing width. This fact is reflected to the kink of the
$\chi^2_{\text{min}}$ contour (thin solid line). If the OPE side
increases, the minimum shifts to finite $\Gamma$ (see next section).
The obtained constraint shows a clear correlation between the mass and the
width as pointed out by us in Ref.~\cite{morita_jpsiprl}. Quantitatively
the relation can vary within $\pm(10-20)$ MeV due to the change of the
temperature dependency. Nevertheless the overall behavior of the
correlation is preserved. Figure \ref{fig:g-m_s350} shows a similar case
but the finite width $\simeq 100$ MeV was chosen as the best fit.
The correlation between the mass shift and the width does not differ
from the other example but the shape of the $\chi^2$ contour slightly
does, reflecting the straight contour of $\chi^2_{\text{min}}$.
In these figures, the area surrounded by the $1\sigma$ or $2\sigma$
correlation lines and one of the $\chi^2$ contour can be regarded as a
possible region of the mass shift and the width including uncertainty
for a given $\sqrt{s_0}$.

\begin{figure}[!tb]
 \includegraphics[width=3.375in]{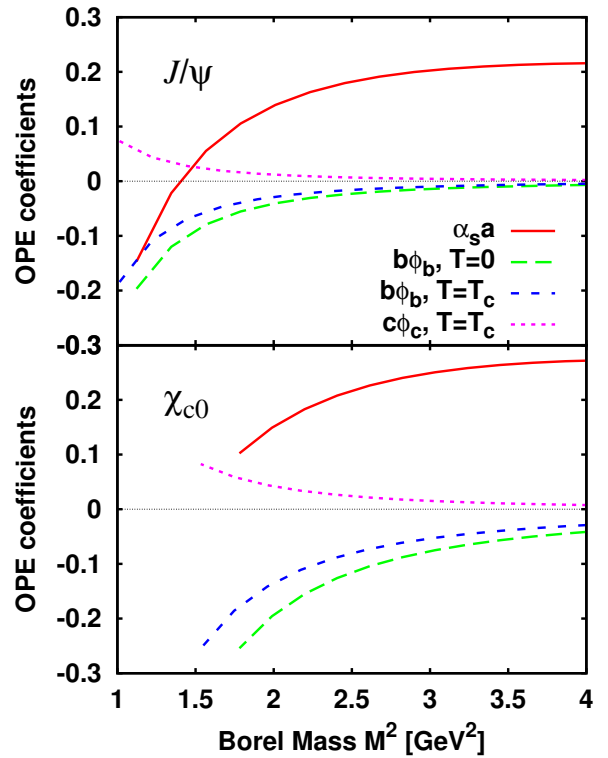}
 \caption{(color online). OPE coefficients of $V$ (upper panel) and $S$
 (lower panel) channels in
 charmonium systems. Solid lines denote the temperature independent
 perturbative correction term $\alpha_s(M^2)a(\nu)$. Long and short
 dashed lines stand for the scalar gluon condensate term
 $b(\nu)\phi_b(T)$ for $T=0$ and $T=T_c$, respectively. Dotted line
 shows the twist-2 contribution $c(\nu)\phi_c(T)$at $T=T_c$.}
 \label{fig:opecof}
\end{figure}

\section{Temperature dependence of spectral parameters}
\label{sec:spectral_result}

\subsection{Implication from OPE side}
\label{subsec_ope_temperature}

Let us begin with examining how temperature dependence of the gluon
condensates is related to the spectral parameters by looking at
the dispersion relation~\eqref{eq:dispersion}. The temperature
dependence of the OPE side [Eq.~\eqref{eq:borel_moment}] comes from the
gluon condensate terms $b\phi_b$ and $c\phi_c$. In
Fig.~\ref{fig:opecof}, we plot the OPE coefficients in
Eq.~\eqref{eq:borel_moment}. One sees that both the scalar gluon
condensate contribution and the twist-2 one increase as temperature increases. These dependencies come
from the fact that $G_0(T)$ and $G_2(T)$ are monotonically decreasing
functions of the temperature while the coefficients $b(\nu)$ and $c(\nu)$
have always negative sign.\footnote{In $P$ channel $b(\nu)$ can be
positive, as understood from Eq.~\eqref{eq:b_ps}, but we found it is
mostly negative for values of $\nu$
corresponding to $M_0^2$ and $\mathcal{M}(M^2)$ retains the property of
increasing function of the temperature.} Therefore, up to dimension four operators,
$\mathcal{M}(M^2)$ in the OPE side is always increasing function of the
temperature.

One also sees that the expansion coefficients of the $P$-wave state are
larger than $S$-wave's at the same Borel mass $M^2$, as also seen in the
moment sum rule case \cite{song09,lee_morita_song_prep}. These properties result in the correct mass splitting between the $J/\psi$ and the $\chi_c$ states, and induce larger mass shift for the $P$-wave states through the derivatives of these
coefficient with respect to $1/M^2$. From the
behavior of the condensate contributions at low $M^2$, one realizes how
the location of the Borel window in the case of charmonium changes with respect to the
temperature. At temperatures close to $T_c$, $M^2_{\text{min}}$ is
determined by the scalar condensate and thus becomes smaller as temperature
increases. It eventually starts to increase when the twist-2
contribution dominates over the scalar one, $|c(\nu)\phi_c| >|b(\nu)\phi_b|$.

>From the dispersion relation \eqref{eq:dispersion}, where the weight of
the integral over the spectral function is positive definite, one obtains the  constraint equation for the
changes of the spectral parameters against the change of the OPE side
discussed above. The phenomenological side
\eqref{eq:phen}--\eqref{eq:phen_cont} has four parameters; effective continuum
threshold $s_0$, pole mass $m$, width $\Gamma$, and overlap $f$.
If only one of these four quantities is allowed to vary as temperature increases, the respective
changes of the parameters needed to match the OPE side are,
\begin{itemize}
 \item $s_0$ : decrease,
 \item $m$ : decrease,
 \item $\Gamma$ : increase,
 \item $f$ : increase.
\end{itemize}
In practice, all of these quantities can change not only to the expected
direction but also to the opposite, as long as the total combined change
of the spectral function matches the OPE side. The Borel transformation
procedure explained in the previous section provides an optimization way
to find out the best set of the changes.

\subsection{Results for charmonia}
\label{subsec:charm}

We carried out the analyses for the charmonia for various temperatures
around $T_c$. For a reference, we summarize the results of $T=0$ in
Table \ref{tbl:charm_T0}. One sees that the masses are well reproduced
by the common parameter set indicated before. Finer tuning on the quark
mass, the coupling constant, and the gluon condensate may improve the
small discrepancies with the experimental data but we are not intending to do
so here since our aim is to investigate the relative change from the vacuum
value induced by the medium.

\begin{table}[tb]
 \caption{Spectral parameters of $c\bar{c}$ systems at
 $T=0$. Experimental masses are taken from Particle data book \cite{pdg2008}.}
 \label{tbl:charm_T0}
 \begin{ruledtabular}
  \begin{tabular}{cccccc}
   System&$\sqrt{s_0}$ [GeV]&$m$ [GeV]&$m_{\text{exp}}$ [GeV]&$M_0^2$ [GeV$^2$]&$f_0$ [GeV$^2$] \\\hline
   $\eta_c$&3.48&2.993&2.980&1.547&0.396 \\
   $J/\psi$&3.54&3.060&3.097&1.971&0.393 \\
   $\chi_{c0}$&3.82&3.406&3.415&2.552&0.303\\
   $\chi_{c1}$&3.78&3.470&3.511&2.810&0.196\\
  \end{tabular}
 \end{ruledtabular}
\end{table}

\begin{figure}[tb]
 \includegraphics[width=3.375in]{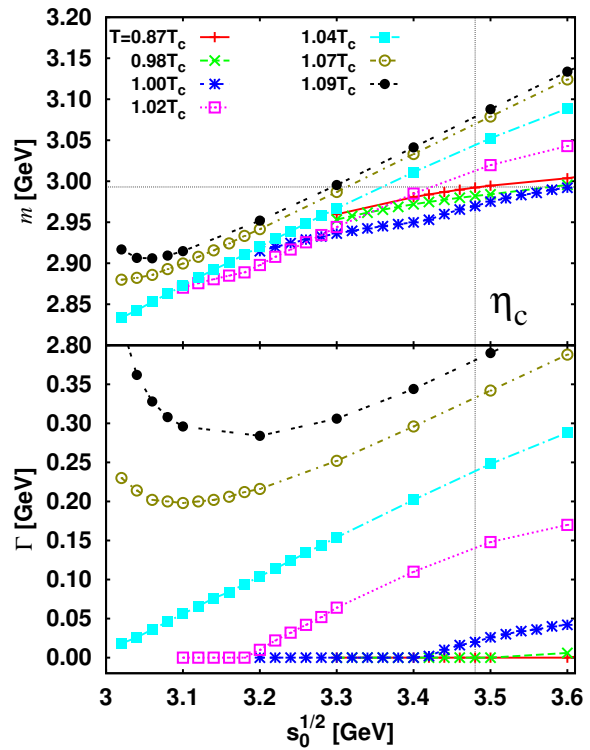}
 \caption{(color online). Constraint on the spectral parameters obtained from the QCD
 sum rule for $\eta_c$. Upper and lower panels show the extracted masses
 and widths as functions of the continuum threshold,
 respectively. Symbols guided by lines represent the different
 temperatures. The vertical and horizontal lines indicate the continuum
 threshold and the mass at $T=0$, respectively.}
 \label{fig:qcdsr_etac}
\end{figure}

\begin{figure}[tb]
 \includegraphics[width=3.375in]{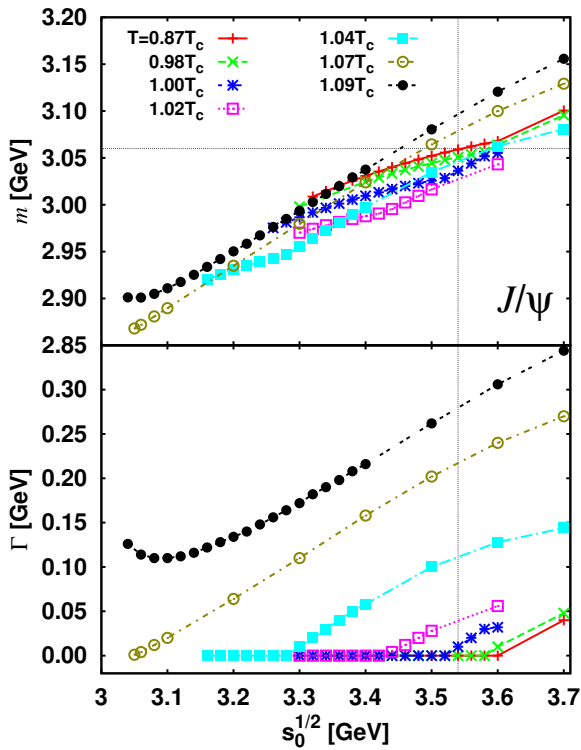}
 \caption{(color online). Same as Fig.~\ref{fig:qcdsr_etac}, but for $J/\psi$.}
 \label{fig:qcdsr_jpsi}
\end{figure}

\begin{figure}[tb]
 \includegraphics[width=3.375in]{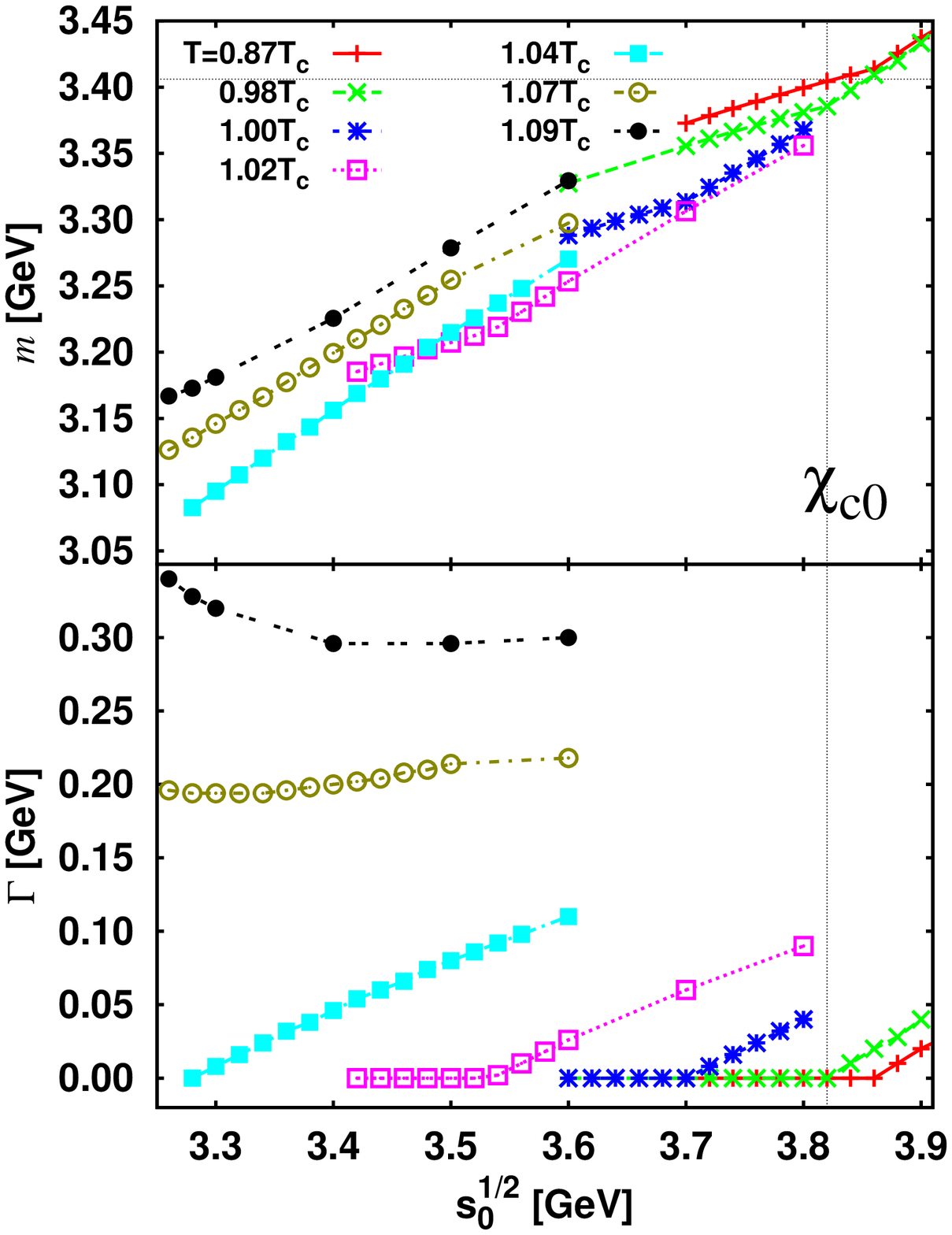}
 \caption{(color online). Same as Fig.~\ref{fig:qcdsr_etac}, but for $\chi_{c0}$.}
\end{figure}

\begin{figure}[tb]
 \includegraphics[width=3.375in]{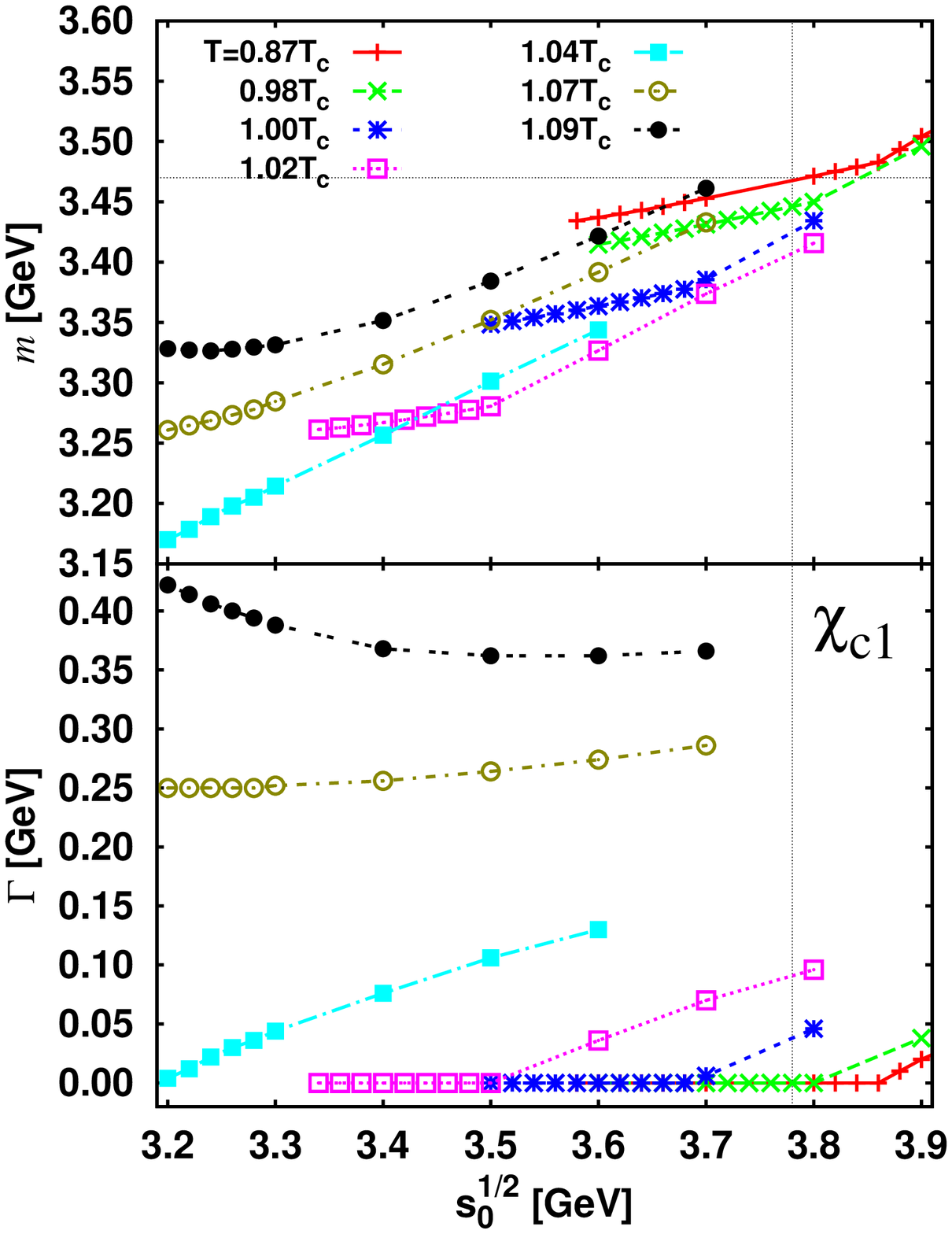}
 \caption{(color online). Same as Fig.~\ref{fig:qcdsr_etac}, but for $\chi_{c1}$.}
 \label{fig:qcdsr_chic1}
\end{figure}

We display the charmonium masses and widths extracted using the method described
in the subsection \ref{subsec:method} as functions of the effective continuum
threshold $\sqrt{s_0}$ in Figs \ref{fig:qcdsr_etac}--\ref{fig:qcdsr_chic1}. One sees the constraints among
$\sqrt{s_0}$, $m$ and $\Gamma$ for various temperatures around $T_c$. Qualitatively all
the four cases are quite similar; the mass increases almost linearly with increasing
$\sqrt{s_0}$ at all the temperatures. While it shows linear increase,
the slope of the mass changes depends on whether the width is zero or
not. One can see
kinks in the mass curves at the same values of the horizontal axis as
those of the width, which vanishes at small $\sqrt{s_0}$.
Although we do not adopt $\sqrt{\chi^2}$ as a way to determine
the best $\sqrt{s_0}$, we notice that it takes the minimum
near the kink among various $\sqrt{s_0}$ values at a fixed
temperature.
At $T=T_{\text{onset}}$, beyond which width must be nonzero to maintain
the Borel stability, the widths shown in the lower panels show linearly
increasing behavior with increasing continuum threshold. The onset temperatures
are summarized in Table \ref{tbl:onsetcharm}. One sees $\eta_c$ starts
to broaden earlier than $J/\psi$ while $T_{\text{onset}}$ of the
$P$-wave states are the same. This may indicate a different temperature effect on the
spin-spin interaction responsible for the mass splitting in the $S$-wave
states. In comparison with what we learned in
Sec.~\ref{subsec_ope_temperature}, one sees the result is more
complicated than the simple analysis by the dispersion relation.
One sees the reduction of the mass always couples with that of the
effective threshold. This is a consequence of the optimization by the
Borel transformation, \textit{i.e.}, using Eq.~\eqref{eq:sumrule} and
looking for the stable curve. To see the reason more explicitly, one can go back to
Fig.~\ref{fig:borelexample}. The reduction of the condensates lowers the
mass without any change in other parameters (see dotted and solid
curves). The requirement of the Borel stability makes the curve flatter
by reducing the effective threshold (see dashed curve). Thus the
downward mass shift always occurs with the reduction of the effective
continuum threshold if there is no broadening.

\begin{figure}[tb]
 \includegraphics[width=3.375in]{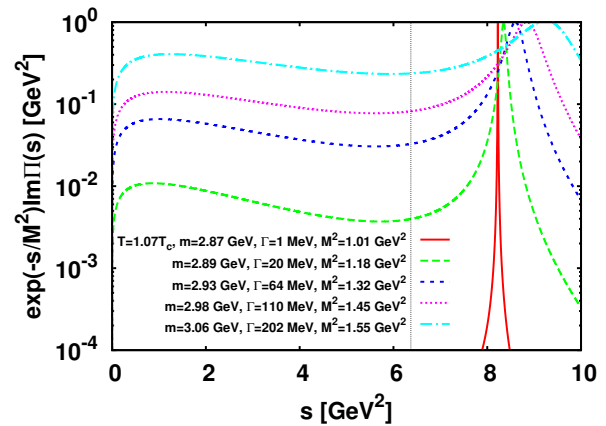}
 \caption{(color online). Integrand of the dispersion integral \eqref{eq:dispersion}
 with the pole term obtained in $T=1.07T_c$. We normalized the different
 lines so that they become unity at $s=m^2$. The thin dotted line
 parallel to the vertical axis indicates the $s=4m_c^2$.}
 \label{fig:bwartifact}
\end{figure}

Behavior of the width at small $\sqrt{s_0}$ and at
$T>T_{\text{onset}}$ seems different from the genuine linear
behavior. It first decreases as $\sqrt{s_0}$ increases, then
turns to increase.
Formally we could obtain stable Borel curves at higher temperatures
than those shown in the figure. At such high temperature, the width
is always at an order of hundreds MeV and the Borel curve is similar to
those displayed in Fig.~\ref{fig:borelexample2}.
We would like to stress, however, that this result may not be a physical
one; in this region, the mass is also small while the width becomes
$100-200$ MeV or more. Clearly the Breit-Wigner ansatz in the
phenomenological side \eqref{eq:phen_pole} which cut off the lower
energy part than $4m_c^2$ does not match with the
dispersion integral \eqref{eq:dispersion}. If we do not impose the
cutoff, the strong suppression factor combined
with the Breit-Wigner form in the Borel-transformed dispersion relation
\eqref{eq:dispersion} leads to numerical artifacts such that the
contribution coming from spectral function much below $4m_c^2$ comprises
a subdominant fraction of the total dispersion integral. We depict an
example taken from $J/\psi$ at $T=1.07T_c$ in Fig.~\ref{fig:bwartifact}. The
same consideration also holds for $\eta_c$, $\chi_{c0}$, and $\chi_{c1}$
at $T \geq 1.07T_c$. One sees that the integral receives large contribution from the  energy region much smaller than
$4m_c^2$ when the width becomes larger, despite the increase of the
mass.
For example, let us consider the change of the width when lowering the
$h\bar{h}$ threshold by 1 GeV$^2$ in the data shown in Fig.~\ref{fig:bwartifact}. With this change, 
$\Gamma=64$ MeV at $\sqrt{s_0}=3.2$ GeV and $\Gamma=110$ MeV at
$\sqrt{s_0}=3.4$ GeV become 30 MeV and 62 MeV, respectively.
This is so because the contribution to the dispersion integral from
$s=4m_c^2-1$ to $s=4m_c^2$ is large enough to compensate the
smaller Breit-Wigner width. We notice, however, that the solution of the
sum rule, Eq.~\eqref{eq:sumrule}, does not exist at near $M_{\text{min}}^2$
for even smaller threshold as in the case shown in
Ref.~\cite{lee_morita_nielsen_exoticwidth}.

We also notice that this artifact is absent in the moment sum
rule up to $n=20$ beyond which it breaks down. Therefore, changing
lower limit of the integration range from $4m_c^2$ to 0 will not affect
the previous results. At present, use of the vacuum dispersion relation,
which cuts off the contribution below $4m_c^2$, seems effective to
estimate the width when its magnitude is less than 100 MeV. To give more quantitative results,
we may need to take into account more detailed structure beyond the
Breit-Wigner ansatz.
Recent resummed perturbative calculation
\cite{burnier09:_heavy} might provide useful information for a better
modeling.
Furthermore, Borel curves at low $M^2$ will be more influenced by
higher dimensional operators we have neglected. Since the width is
sensitive to the low $M^2$ behavior of the Borel curves, as shown in
subsection \ref{subsec:method}, it may receive sizable correction from
those operators. At present, temperature dependence of the higher
dimensional operators is poorly known. More quantitative
analysis of the width in the non-perturbative manner thus needs further efforts.

\begin{table}[tb]
 \caption{Onset temperatures of the width $T_{\text{onset}}$ for charmonia}
 \label{tbl:onsetcharm}
 \begin{ruledtabular}
  \begin{tabular}{ccccc}
   &$\eta_c$&$J/\psi$&$\chi_{c0}$&$\chi_{c1}$ \\\hline
   $T_{\text{onset}}$&1.04&1.07&1.05&1.05
  \end{tabular}
 \end{ruledtabular}
\end{table}

\begin{figure}[tb]
 \includegraphics[width=3.375in]{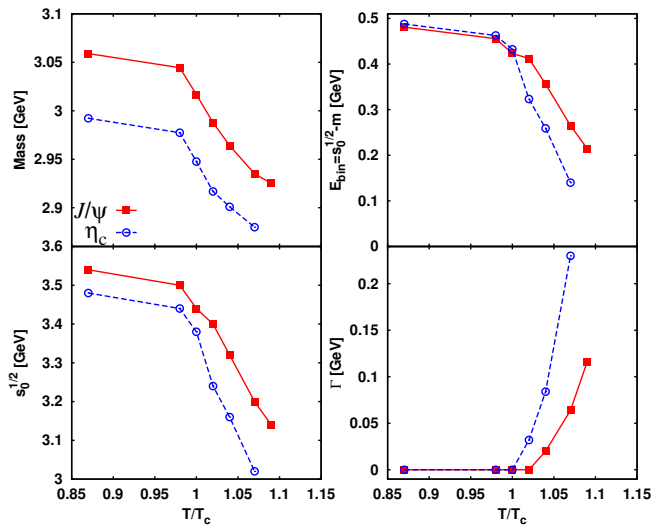}
 \caption{(color online). Temperature dependence of the spectral
 parameters of $J/\psi$ (closed symbol) and $\eta_c$ (open symbol)
 extracted from QCD sum rules combined with the second order Stark
 effect. The masses (top-left panel) are obtained from the second order
 Stark effect and used as inputs for the sum rule analyses. Other
 quantities are obtained by reading off the result of
 Figs.~\ref{fig:qcdsr_etac} and \ref{fig:qcdsr_jpsi} which match with
 the mass shifts.}
 \label{fig:starkcharm}
\end{figure}

Since Figs. \ref{fig:qcdsr_etac}--\ref{fig:qcdsr_chic1} give only
constraints, one needs to specify one of those
spectral parameters to discuss specific temperature dependencies of each parameter.
Previous analyses
\cite{morita_jpsiprl,morita_jpsifull,song09,lee_morita_song_prep}
correspond to $s_0\rightarrow \infty$ limit. For instance, if $\sqrt{s_0}$
retains the vacuum value, the mass decreases rapidly
until $T=T_c$ in $\eta_c$ and $T=1.02T_c$ in $J/\psi$.
We do not show the results of constant $\sqrt{s_0}$ for the $P$-wave
states, as Eq.~\eqref{eq:sumrule} has no solution at certain region
inside the Borel window and thus $\chi^2$ [Eq.~\eqref{eq:chi2}]
cannot be evaluated before it reaches the minimum as a function of the
width.
This absence of the solution actually occurs in $S$-wave
cases also, especially at larger $\sqrt{s_0}$
and comes from the non-monotonic behavior of Borel transformation of the Breit-Wigner function
\cite{lee_morita_nielsen_exoticwidth}. Note that this does not mean the
corresponding parameter sets are completely excluded, since one may
choose another (narrower in most cases) Borel window such
that the solution exists in any $M^2 \in [M^2_{\text{min}},M^2_{\text{max}}]$.
Nevertheless, from the almost linear dependence of the
mass and width on $\sqrt{s_0}$, one can extrapolate the lines up to the
desired value to have a rough estimate.  Then, one finds in all the
channels that the mass first decreases then the width starts to increase, as the temperature increases when $\sqrt{s_0}$ is
held fixed.  Note that this transition of the temperature dependence of the mass
is caused by the start of the broadening of its width. We would like to point out
that the analysis at $s_0=\text{constant}$ is not the same as that at
$s_0\rightarrow \infty$ in which the determination of the flattest mass
curve by $\chi^2$ does not make sense. When the mass is held to its vacuum
value, the constraint is satisfied by the increase in both $\sqrt{s_0}$
and the width. For example, in the $J/\psi$ case,
$\sqrt{s_0}$ becomes 3.6 GeV and $\Gamma=128$ MeV at $T=1.04T_c$ while
$\Gamma=0$ is still possible if the mass and $\sqrt{s_0}$
decrease to 2.92 GeV and 3.16 GeV, respectively.

At present, the temperature dependence of the continuum threshold is not
clearly known yet. In fact, our threshold parameter should be regarded
as an effective one since we do not take radial excited states such as
$\psi'(2S)$ and $\eta_c(2S)$ into account nor temperature dependent
behaviors near the threshold \cite{cabrera07:_t,mocsy07,laine07:_qcd}.
Nevertheless, if those states dissociate
at as low temperatures as $T_c$, one may regard $\sqrt{s_0}$ as a
physical threshold within temperatures between $T_c$ and a certain temperature at which
the pole position becomes so close to the threshold that the pole and the
continuum part of the model spectral function starts to overlap.
For the $P$-wave states, the model will be better due to the absence of
excited states below the threshold.
One might be able to interpret the asymptotic value of the
quark-antiquark potential as
the continuum threshold \cite{mocsy06}, which decreases as temperature
increases irrespective of the choice of the potential
\cite{Karsch04,Wong05}. This fact might be related to the decrease of
the mass and subsequent dissolution of the $D$ mesons~\cite{Dominguez}.
In this case, the obtained constraints give the downward mass shifts in
the all channels. If the reduction is strong, only the mass shift occurs
without broadening up to $T=T_{\text{onset}}$. If not, the widths will start
to broaden gradually together with the moderate downward shift of the mass.

Another external constraint can be obtained from the second order Stark
effect in QCD \cite{peskin79,luke92,lee03,lee_morita_stark}. Although the
applicability to the charmonium systems is marginal, it gives a genuine
downward mass shift due to the rapid increase of the color electric
condensate \cite{lee_morita_stark}. For illustration in the case of
downward mass shift, we combined the result of
Ref.~\cite{lee_morita_stark} with those of Figs.~\ref{fig:qcdsr_etac}
and \ref{fig:qcdsr_jpsi},
by finding the masses in Figs.~\ref{fig:qcdsr_etac} and
\ref{fig:qcdsr_jpsi} that matches with the results of the second
order Stark effect and then looking at the corresponding continuum
threshold and width. The results of the masses, continuum thresholds,
binding energies defined by $E_{\text{bin}}=\sqrt{s_0}-m$, and the
widths of $J/\psi$ and $\eta_c$ are displayed in
Fig.~\ref{fig:starkcharm}.
As explained, the results for $T > T_{\text{onset}}$ are marginal.
Moreover, the second order Stark effect has also limitation of
applicability at this temperature region as the change of the electric condensate value becomes too large. Indeed the mass obtained from
the Stark effect at $T > 1.09T_c$ becomes smaller than the smallest mass
in Figs.~\ref{fig:qcdsr_etac} and \ref{fig:qcdsr_jpsi}, indicating the
breakdown of the OPE in the Stark effect. Hence, we emphasize that any
extrapolation of Fig.~\ref{fig:starkcharm} to higher temperature is not
appropriate. Apart from the marginal region, one sees that the downward
mass shift smaller than the maximum given by QCD sum rules, as already
found in Ref.~\cite{lee_morita_stark}, leads to broadening just above
$T_c$.
These temperatures, $1.02T_c$ for $\eta_c$ and $1.04T_c$ for $J/\psi$, are
lower than the corresponding onset temperatures. One also sees that the continuum
thresholds suddenly decrease around $T_c$ as in the case for the masses.
It is quite intriguing to see that similar results are obtained in the potential model approaches, which utilizes the confinement force that can not be derived within the OPE formalism.  Since the continuum thresholds change more rapidly,
the resultant binding energy also drastically decreases across
$T_c$. At the marginal temperatures, $E_{\text{bin}}$ is still around
100--200 MeV but the widths also become sizable due to thermal activation by
gluons. We cannot draw conclusion on the dissociation of
the charmonia from these results, since $\Gamma > 100$ MeV will have to be
examined more carefully by incorporating higher dimensional operators
and more realistic spectral function. Below
$T_{\text{onset}}$, one finds $E_{\text{bin}}$ is still larger than
$\Gamma/2$, indicating binding just above $T_c$.
Furthermore, one does not see any broadening below $T_c$. That is
in line with our previous finding in Ref.~\cite{lee_morita_stark}, where
the effect of the continuum was not taken into account.
Finally we would like to stress that all the
spectral parameters show sudden change across $T_c$, as shown in
Fig.~\ref{fig:starkcharm}, reflecting the
abrupt change of the gluon condensates at this temperature and thus the
QCD phase transition.
Moreover, as discussed before, even if one of these parameters remains constant and retains its vacuum value, the QCD sum rule constraints force other quantities to exhibit such critical behaviors.

\subsection{Results for bottomonia}

\begin{table}[tb]
 \caption{Spectral parameters of $b\bar{b}$ systems at $T=0$. Experimental masses are taken from Particle data book \cite{pdg2008}.}
 \label{tbl:bottom_T0}
 \begin{ruledtabular}
  \begin{tabular}{cccccc}
   System&$\sqrt{s_0}$ [GeV]&$m$ [GeV]&$m_{\text{exp}}$ [GeV]&$M_0^2$ [GeV$^2$]&$f_0$ [GeV$^2$] \\\hline
   $\eta_b$&10.28&9.392&9.389&12.31&2.199\\
   $\Upsilon$&10.34&9.447&9.460&13.68&2.034 \\
   $\chi_{b0}$&10.73&9.949&9.859&13.08&0.8\\
   $\chi_{b1}$&10.34&10.09&9.893&14.13&0.492\\
  \end{tabular}
 \end{ruledtabular}
\end{table}

\begin{figure}[t]
 \includegraphics[width=3.375in]{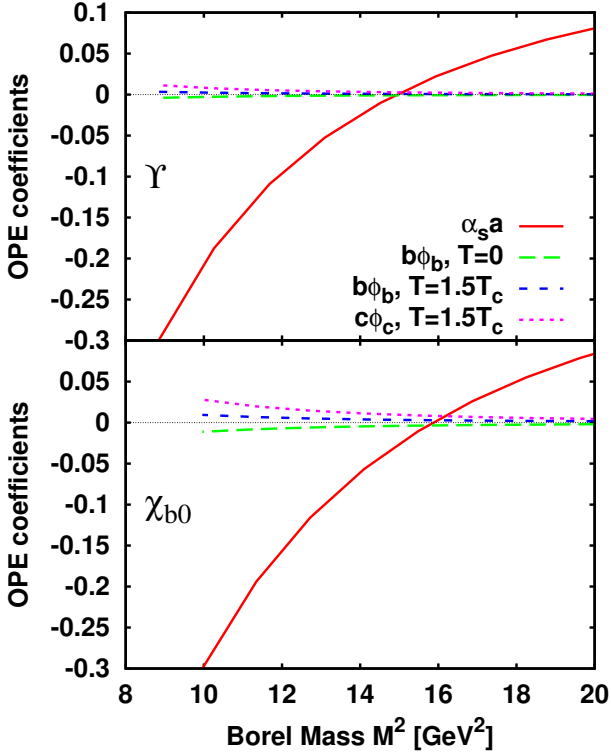}
 \caption{(color online). Same as Fig.~\ref{fig:opecof}, but for $b\bar{b}$ systems}
 \label{fig:ope_bbbar}
\end{figure}

\begin{figure}[!htb]
 \includegraphics[width=3.375in]{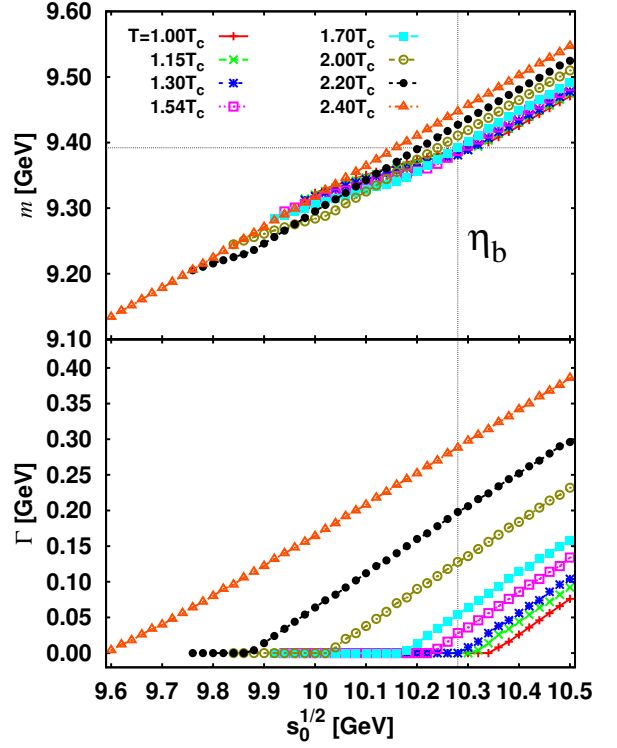}
 \caption{(color online). Same as Fig.~\ref{fig:qcdsr_etac}, but for $\eta_b$.}
 \label{fig:qcdsr_etab}
\end{figure}

\begin{figure}[!htb]
 \includegraphics[width=3.375in]{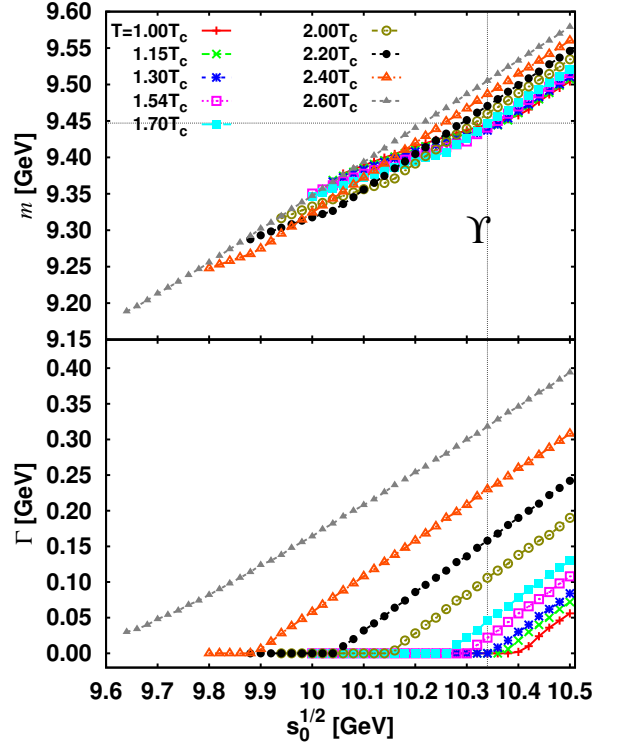}
 \caption{(color online). Same as Fig.~\ref{fig:qcdsr_etac}, but for $\Upsilon$.}
\end{figure}

\begin{figure}[!htb]
 \includegraphics[width=3.375in]{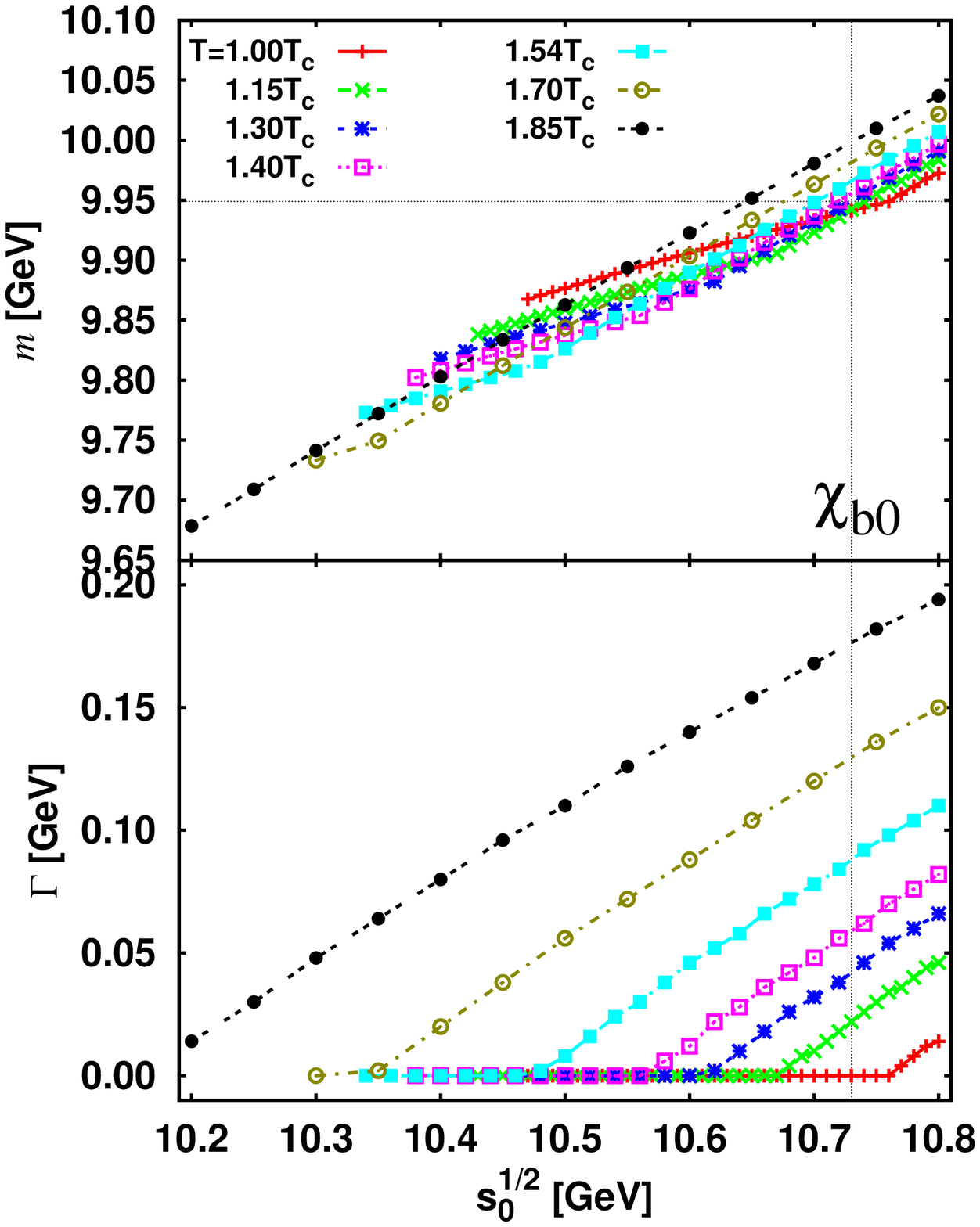}
 \caption{(color online). Same as Fig.~\ref{fig:qcdsr_etac}, but for $\chi_{b0}$.}
\end{figure}

\begin{figure}[htb]
 \includegraphics[width=3.375in]{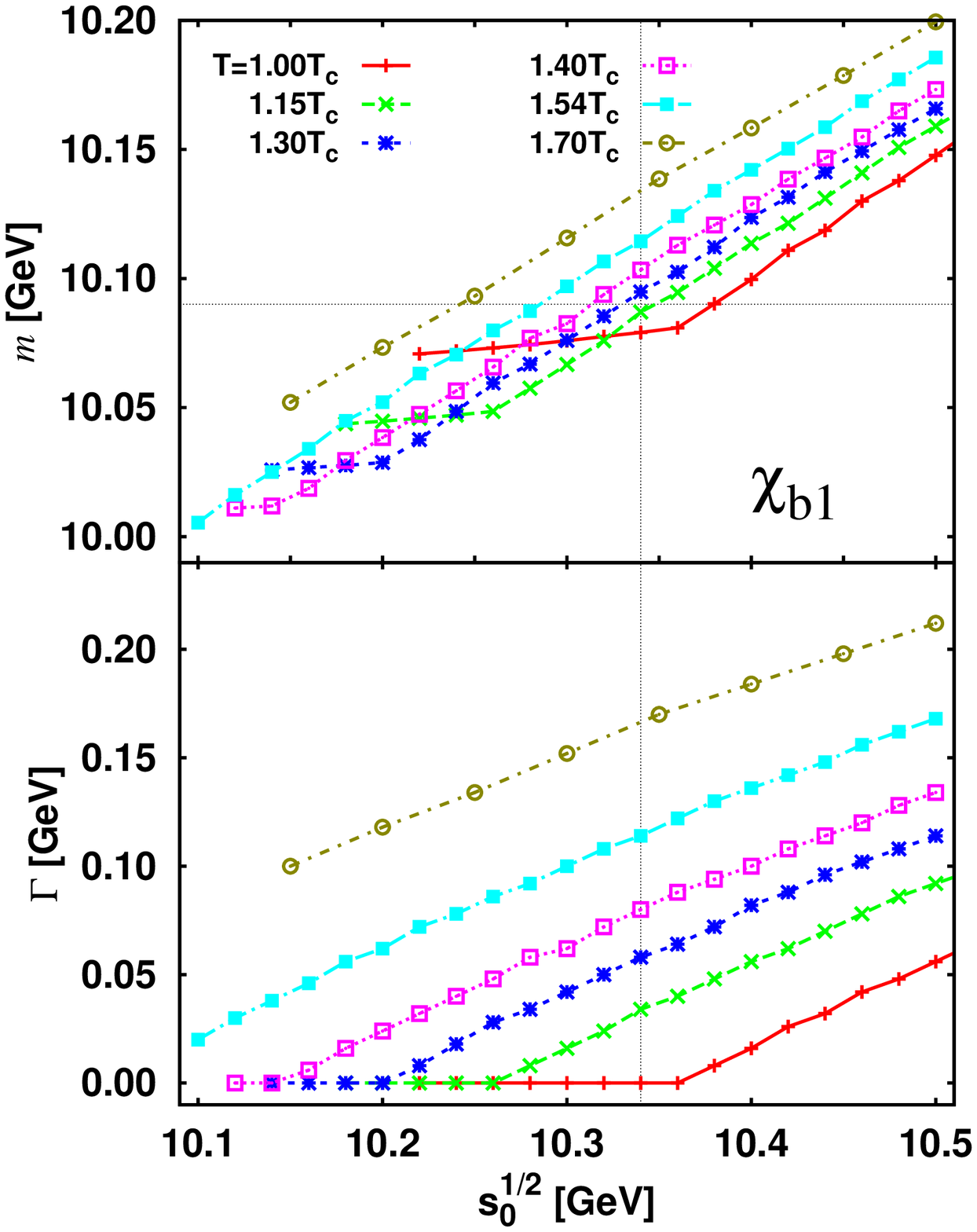}
 \caption{(color online). Same as Fig.~\ref{fig:qcdsr_etac}, but for $\chi_{b1}$.}
 \label{fig:qcdsr_chib1}
\end{figure}

\begin{table}[tb]
 \caption{Onset temperatures of the width $T_{\text{onset}}$ for bottomonia}
 \label{tbl:onsetbottom}
 \begin{ruledtabular}
  \begin{tabular}{ccccc}
   &$\eta_b$&$\Upsilon$&$\chi_{b0}$&$\chi_{b1}$ \\\hline
   $T_{\text{onset}}$&2.40&2.56&1.87&1.50
  \end{tabular}
 \end{ruledtabular}
\end{table}

\begin{figure}[!htb]
 \includegraphics[width=3.375in]{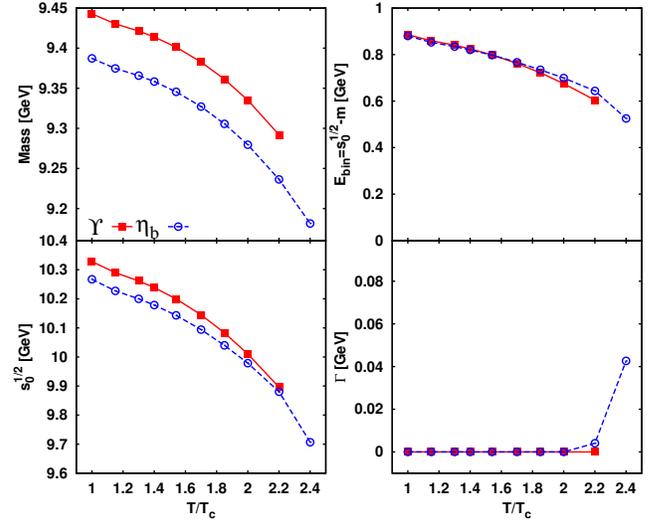}
 \caption{(color online). Temperature dependence of the spectral
 parameters of $\Upsilon$ (closed symbols) and $\eta_b$ (open symbols)
 extracted from QCD sum rules combined with the second order Stark effect.}
 \label{fig:stark_bottom}
\end{figure}

We also calculate the in-medium changes of the spectral property of the
bottomonia using the same framework. Results for $T=0$
are summarized in Table \ref{tbl:bottom_T0}. As in the charmonium case, the sum rule works well
for the bottomonium masses in the vacuum. In the case of bottomonium,
the relative contribution of the dimension four operator to the OPE is
much smaller than that of the charmonium because of the $m_h^{-4}$
dependence in the Wilson coefficient as seen in Eqs.~\eqref{eq:phib} and
\eqref{eq:phic}. Hence, its spectral property is
much less affected by the change of the gluon condensates coming from
the temperature effects. This fact allows us to go to much higher
temperatures than in the charmonium cases until the dimension four
contributions become so large as to break the Borel stabilities. We
display some of the OPE coefficients at $T=0$ and $T=1.5T_c$ in
Fig.~\ref{fig:ope_bbbar}.
One sees that the power correction terms are much smaller than the
leading perturbative correction term and might in fact be similar in
magnitude to that of the next higher order radiative correction
\cite{chetyrkin97}.
Therefore, while the separation scale in the present case is large
enough for the OPE to provide a qualitatively reliable guide, further
efforts are needed to obtain a quantitatively accurate estimate of the
spectral property.
>From Fig.~\ref{fig:ope_bbbar}, one also sees that the twist-2 term
dominates the temperature effect at $1.5T_c$ and also at temperatures
where the spectral modification becomes sizable as will be seen
below. Since the perturbative effects are more dominant in the present
case than in the charmonium cases, a detailed comparison with the
resummed perturbative approach \cite{burnier09:_heavy} might be useful
to understand the interplay between the  perturbative and the
non-perturbative effects at these temperatures.

We show the constraints among the effective continuum threshold, mass
and width for each of the  bottomonium states in
Figs.~\ref{fig:qcdsr_etab}--\ref{fig:qcdsr_chib1}.
The basic features are the same as in the charmonium cases, except now
the sudden change across $T_c$ has disappeared.
As the position of the mass is far from $2m_b=8.24$ GeV, the
unphysical behavior of the width seen at $T > T_{\text{onset}}$ in the
charmonium cases is absent in the bottomonium cases. The onset
temperatures are summarized in Table \ref{tbl:onsetbottom}. One sees that the $S$-wave
states have the narrow pole solution up to $T\sim 2.4T_c$ and $P$-waves do so
up to $1.5T_c$, suggesting survival of these states up to somewhat
higher temperature than those suggested by a potential model analysis \cite{mocsy08}.
The maximum mass shifts are obtained near the onset temperatures and are
found to be around 200 MeV for all the channels except for the axial
vector.
The mass shift of the $P$-wave states is found to be twice as large as
that of the $S$-wave states at a fixed temperature as in the
charmonium states. Unlike the charmonium, one sees a significant
difference in $T_{\text{onset}}$ between the $S$ and $A$ channel.
This is a manifestation of the dependence of $T_{\text{onset}}$ on the
Borel window that was discussed in Sec.~\ref{subsec:method}. In this
case, the additional condition, the perturbative correction less than
0.3 leads to a significant difference of $M_{\text{min}}^2$ between $S$
and $A$ channel such that the Borel window of $A$ channel becomes much
narrower. If one set a common $M_{\text{min}}^2$ for instance,
$T_{\text{onset}}$ does not differ so much. Indeed the Borel stability
in the $\chi_{b}$ states is more marginal than other cases, as indicated
in Ref.~\cite{Reinders_NPB186} as the difficulty of establishing the
``plateau'' in the moment sum rule. For the $A$ channel in
the present case, even at $T=0$, $M^2_{\text{min}}$ is larger than
$M_0^2$ at $\sqrt{s_0}=\infty$. Although one can obtain the stability by
relaxing the criterion for the $M^2_{\text{min}}$, the result would be
less reliable since the $M^2_{\text{min}}$ is so chosen as to validate
the perturbative expansion. After all, while we could obtain reasonable
description of $\chi_b$ states at $T=0$ and plausible in-medium changes
of them, it has an intrisic ambiguity in the quantitative results.

For the $S$-wave states, we also extract the results from combining the
constraints with the second order Stark effect which are expected to be
more reliable in the
bottomonium systems. Figure \ref{fig:stark_bottom} shows the results for
the mass, the effective continuum threshold, the binding energy, and
width of $\Upsilon$
and $\eta_b$. One sees that the changes as a function of the temperature
are rather moderate;
this reflects the smaller effects from the gluon condensates.
Especially there is no significant brodening in both channels up to
$T=2.2T_c$. One should note, however, that the second order Stark effect
gives the larger mass shifts than that of maximum given by the QCD sum
rule at $T =2.4T_c$ in $V$ channel. This is similar to what happened at
$T > 1.09T_c$ in the case of $J/\psi$ (See Sec.~\ref{subsec:charm}).
The heavier quark mass enables us to extend the OPE to higher temperature,
but it seems to break down at   $T=2.4T_c$.
Since the $P$ channel exhibits larger spectral change than the $V$ channel
as in the charmonium cases, the maximum mass shift of $\eta_b$ given by
the QCD sum rule is always smaller than that from the second order Stark
effect. One sees small broadening at $T \geq 2.2T_c$.

\section{Imaginary time correlators}
\label{sec:IMC}

The spectral parameters obtained from QCD sum rules at finite
temperature have shown sizable modifications from the vacuum values.
To confirm the findings, it is desirable to compare the results with the first principle lattice calculation. Unfortunately the direct evaluation of the
spectral function of the heavy quarkonia through MEM has insufficient
resolution to identify the spectral changes of order of 100 MeV. In this
section, we will construct
model spectral functions at finite temperatures as well as in the vacuum using the previously obtained QCD sum rule results. Then, we reconstruct
the imaginary time correlators  via the dispersion relation, discuss how the spectral modification affects the correlator, and then compare them with the  lattice results which are  more accurately calculated.

\subsection{Relation of spectral function with the imaginary time correlator}

The imaginary time correlator $G(\tau,T)$ is related to the spectral
function via the dispersion relation
\begin{equation}
 G(\tau,T) = \int_{0}^{\infty}d\omega K(\omega,\tau;T)\rho(\omega,T)\label{eq:imc}
\end{equation}
where the integration kernel $K(\omega,\tau;T)$ is
\begin{equation}
 K(\omega,\tau;T) = \frac{\cosh[\omega(\tau-1/2T)]}{\sinh (\omega/2T)},\label{eq:kernel}
\end{equation}
of which the zero temperature limit is $e^{-\omega \tau}$.
To see the temperature effect on the spectral function, one usually
computes the ratio of this
correlator to the reconstructed one $G(\tau,T)/G_{\text{rec}}(\tau,T)$
with $G_{\text{rec}}$ defined as
\begin{equation}
 G_{\text{rec}}(\tau,T)= \int_{0}^{\infty} d\omega \, \rho(\omega,T=0)K(\omega,\tau;T)
\end{equation}
which has temperature dependence coming only from the kernel.

We construct a model spectral function to be put into Eq.~\eqref{eq:imc}
from the phenomenological side \eqref{eq:phen}--\eqref{eq:phen_pole}
\begin{align}
 \rho^{\text{pc}}(\omega)&= \frac{C_J\omega^2}{\pi}\left[
 \text{Im}\tilde{\Pi}^{\text{pole}}(\omega^2) +
 \text{Im}\tilde{\Pi}^{\text{cont}}(\omega^2)\right]\label{eq:imc_rho_pc}
\end{align}
with $C_J=1$ for $P$ and $S$ channels and 3 for $V$ and $A$ channels. The subscript ``pc'' denotes the ``pole+continuum''.
We relate them to the spatial components of the spectral function
for $V$ and $A$ channel, in order to compare them with lattice
calculation. In  $A$ channel, although lattice calculation uses an axial
vector current of $j_\mu = \bar{h}\gamma_\mu \gamma_5 h$ while we use
the conserved part $J^\mu = \eta^{\mu\nu}j_\nu$ by multiplying $\eta_{\alpha\beta}=(q_\alpha
q_\beta/q^2-g_{\alpha\beta})$, the above expression still holds.

Putting each part of the model spectral function into Eq.~\eqref{eq:imc},
one obtains the following formulae;
\begin{align}
 G^{\text{pole}}(\tau,T)&= \begin{cases}
		    \displaystyle
		    \frac{C_Jmf_0}{2\pi}\frac{\cosh[m(\tau-1/2T)]}{\sinh(m/2T)}&
		    \Gamma=0\\
			    \displaystyle \frac{C_J f\Gamma}{\pi} \int_{0}^{\infty}
		    \frac{\omega^3d\omega}{(\omega^2-m^2)^2+\omega^2
		    \Gamma^2} \\
			\displaystyle    \times \frac{\cosh[\omega(\tau-1/2T)]}{\sinh(\omega/2T)} &
		    \Gamma \neq 0
		   \end{cases}\\
 G^{\text{cont}}(\tau.T)&= \frac{C_J}{\pi}\int_{\sqrt{s_0}}^{\infty}
 d\omega \, \omega^2 \text{Im}\tilde{\Pi}^{J,\text{pert}}(\omega^2)
 \frac{\cosh[\omega(\tau-1/2T)]}{\sinh(\omega/2T)}.
\end{align}
As for the peak strength parameter $f$ and $f_0$, One can obtain it by
using the dispersion relation \eqref{eq:dispersion} after determining the
other three parameters as
\begin{align}
 f_0&=
 e^{m^2/M_0^2}[\mathcal{M}(M_0^2)-\mathcal{M}^{\text{cont}}(M_0^2)],\\
 f &=
 \frac{\mathcal{M}(M_0^2)-\mathcal{M}^{\text{cont}}(M_0^2)}{\displaystyle
 \Gamma
 \int_{4m_h^2}^{\infty} ds \, e^{-s/M_0^2} \frac{\sqrt{s}}{(s-m^2)^2+s\Gamma^2}}.
\end{align}
We adopt the value of the Borel mass at $M^2=M_0^2$ where the
property of the pole part was determined. This
choice, however, can result in an unphysical behavior of the reconstructed
correlator ratio $G/G_{\text{rec}}$ near $\tau\simeq 0$ due to the
sensitivity to the high energy part of the specral function.
The continuum part in the $P$ and the $S$ channel has a singular
behavior in the high energy
limit \cite{Reinders_NPB186} such that a slight deviation in $\alpha_s$
can lead to a sizable difference in the imaginary time correlator near
$\tau=0$. While this does not matter at $\tau \geq 0.1$ fm and in the
QCD sum rule analysis due to the large suppression of the high energy
part by the Borel transformation, we use the same the value of
$\alpha_s$ at finite temperarature as that of $T=0$ by fixing $M_0^2$
so that $G/G_{\text{rec}}\rightarrow 1$ as $\tau\rightarrow 0$ at any temperature.

Explicit temperature, not divided by
$T_c$, need to be specified in the kernel. While our gluon condensates
have been taken from
the lattice calculation with $T_c=264$ MeV \cite{Boyd_NPB469}, we normalize the temperature dependence in the imaginary time correlator calculation to $T_c=295$ MeV, which corresponds to the normalization used in the lattice calculation that  we will be comparing our results to \cite{jakovac07}.

It has been emphasized that a peak of the spectral function at
$\omega=0$ gives a constant contribution to the imaginary time correlator \cite{umeda07}.
Although we have ignored this contribution in the QCD sum rules, as
explained above, this term is necessary for proper comparison of $G(\tau,T)$.
Here, we adopt the expression calculated for free heavy quarks which is
proportional to $\omega \delta(\omega)$. In this case, the spectral
functions have been calculated and given in
\cite{aarts05}.\footnote{There is a misprint in Ref.~\cite{aarts05}
pointed out in Ref.~\cite{mocsy06}.} The zero mode
(scattering) parts for $V,P,S$ and $A$ channels are given by
\begin{equation}
 \rho^{\text{scat}}(\omega) = N_c \omega \delta(\omega)(c_1I_1-c_2I_2).
\end{equation}
The numerical constants $c_1$ and $c_2$ are summarized in Table
\ref{tbl:zeromode} and

\begin{table}[tb]
 \caption{Numerical constants in zero mode spectral function of various channels.}
 \label{tbl:zeromode}
 \begin{ruledtabular}
  \begin{tabular}{ccccccc}
   &$P$&$V(\rho^{ii})$&$V(\rho^{\mu}_\mu)$&$S$&$A(\rho^{ii})$&$A(\rho^{\mu}_{\mu})$
   \\\hline
   $c_1$&0&0&$-2$&2&6&6 \\
   $c_2$&0&2&2&$-2$&$-4$&$-6$ \\
  \end{tabular}
 \end{ruledtabular}
\end{table}
\begin{align}
 I_1 &= -2\int \frac{d^3 \boldsymbol{k}}{(2\pi)^3}\frac{dn_k}{d\omega_k}\nonumber\\
 I_2 &= -2\int \frac{d^3 \boldsymbol{k}}{(2\pi)^3}\frac{dn_k}{d\omega_k}
  \frac{\boldsymbol{k}^2}{\omega_k^2}\label{eq:I1I2}
\end{align}
where $n_k = (e^{\omega_k/T}+1)^{-1}$ and
$\omega_k = \sqrt{\boldsymbol{k}^2+m_h^2}$.
Putting these expressions into Eq.~\eqref{eq:imc}, finally one obtains the constant
contribution to the imaginary time correlator
\begin{equation}
 G^{\text{scat}}(\tau,T)= N_cT(c_1I_1-c_2I_2).\label{eq:imczeromode}
\end{equation}
Hereafter, we adopt the three component model
$G(\tau,T)=G^{\text{pole}}+G^{\text{cont}}+G^{\text{scat}}$
with spectral parameters taken from the
results of the QCD sum rule as our model imaginary time correlator which
we compare with the lattice QCD result shown in Ref.~\cite{jakovac07}.

\begin{figure}[tb]
 \includegraphics[width=3.375in]{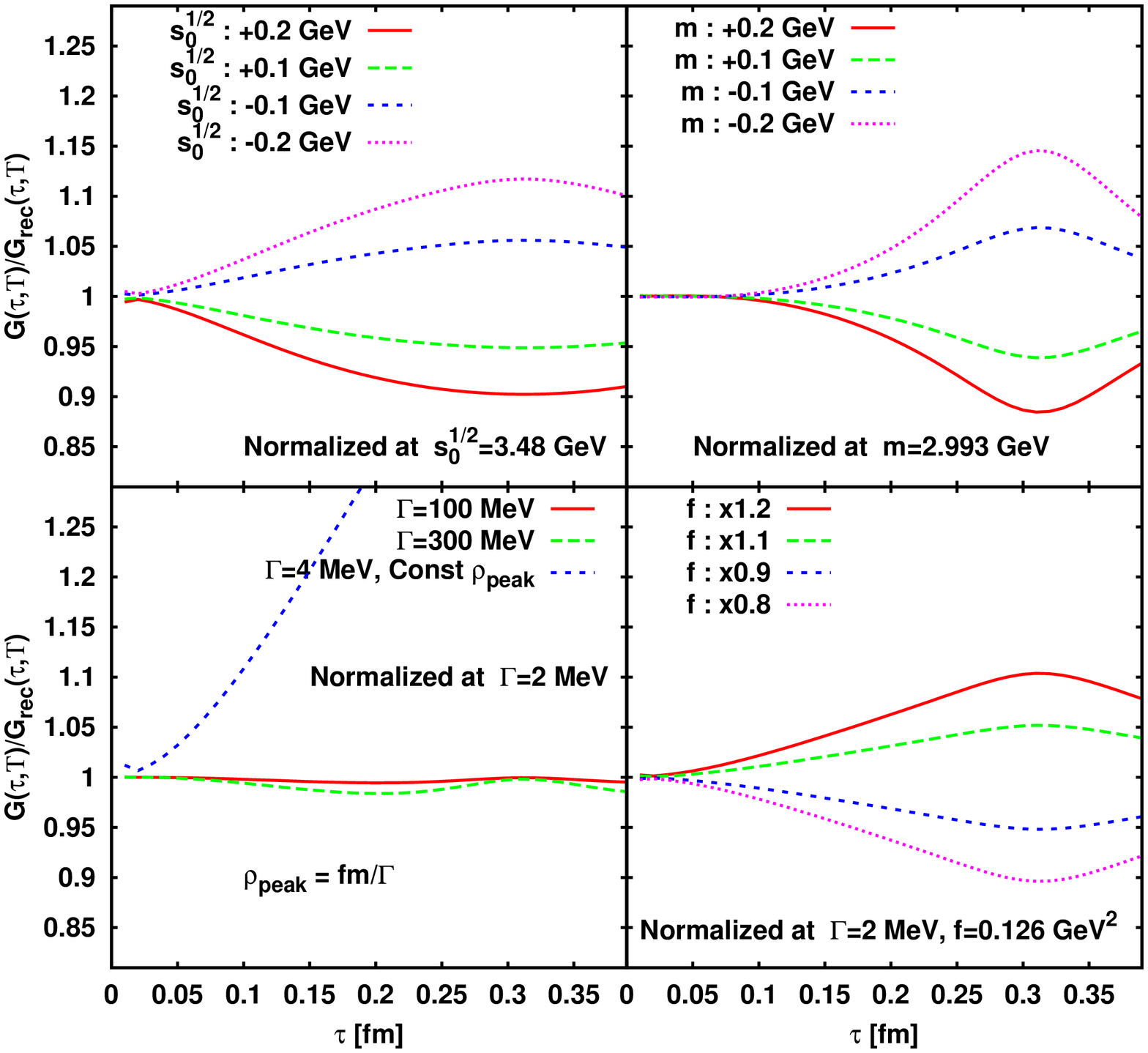}
 \caption{(color online). $G/G_{\text{rec}}$ in several cases of modification of
 \textit{single} spectral property. Upper left : continuum threshold,
 upper right : mass, lower left : width, and lower right : peak
 height. The zero temperature spectral function in $G_{\text{rec}}$ is
 taken from $\eta_c$ case thus constant contribution is absent. For
 the lower panel, we utilized $T=0$ spectral parameters obtained by
 putting $\Gamma=2$ MeV for comparison.}
 \label{fig:imc_example}
\end{figure}

Before proceeding, it is useful to see how the typical spectral
changes seen in the QCD sum rule affect the ratio of the imaginary time
correlator $G/G_{\text{rec}}$. In Fig.~\ref{fig:imc_example}, we plot several examples in which only one of the four spectral parameters is changed. Similar to what was shown in subsection \ref{subsec_ope_temperature}, one can
understand the qualitative behavior from the dispersion relation for the
imaginary time correlator \eqref{eq:imc} due to the positivity of the
kernel.
When the modification of the spectral parameters increases the spectral
sum, $G/G_{\text{rec}}$ also increases. One sees that a rather small
modifications of the continuum threshold and the mass, of order of 100 MeV,
lead to more than 10\% change in $G/G_\text{rec}$, whereas
less than 3\% changes have been observed in the lattice
QCD calculation of the $P$ channel \cite{jakovac07}. On the other hand, the change of width does not
affect $G/G_{\text{rec}}$, as seen in the lower-right panel. This is so
because the increase of width implies reduction of the peak height
$\rho(\omega^2=m^2)=fm/\Gamma$. If one artificially tries to preserve the height by increasing $f$ as
well as $\Gamma$, even small increase of the width makes
$G/G_{\text{rec}}$ increase very quickly, as shown in the
dotted line in the figure.

\subsection{Comparison with lattice QCD results}

\subsubsection{Charmonium}

\begin{figure}[tb]
 \includegraphics[width=3.375in]{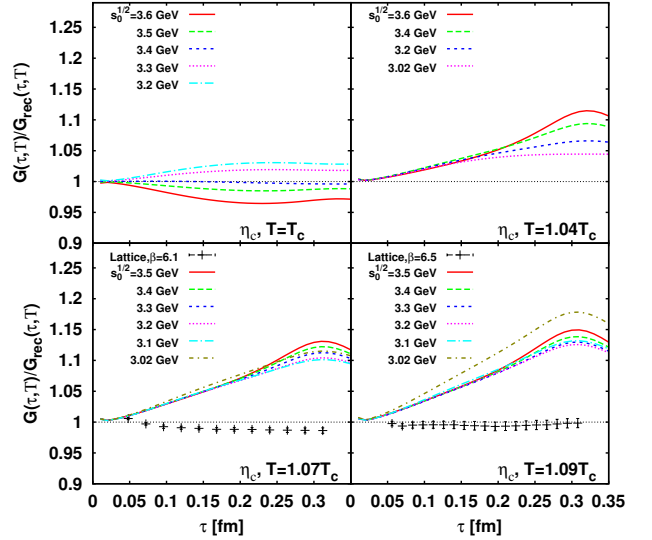}
 \caption{(color online). Imaginary time correlators for $\eta_c$. Each
 of panels shows different temperature cases, $T=T_c$, $1.04T_c$, $1.07T_c$, and
 $1.09T_c$. For $T=1.07T_c$ and $T=1.09T_c$, we plot results from
 lattice QCD shown in Ref.~\cite{jakovac07} with crosses.}
 \label{fig:imc_etac}
\end{figure}

We start with the charmonium in the  $P$ channel ($\eta_c$) to which no zero mode
contributes. Unfortunately, the available lattice data are for
$T=0.87T_c$, $1.07T_c$, $1.09T_c$ and so on while our sum rule results
of interest are just around $T_c$. At $0.87T_c$, spectral change is
almost negligible because of the tiny change of the gluon condensates.
At the next lowest temperature, $1.07T_c$, it is already above the onset
temperature therefore our result for the charmonium is quantitatively
obscure. Between the lowest and the next lowest temperature, however,
$G/G_{\text{rec}}$ on the lattice seems unchanged because it is almost
unity at both temperatures. Sizable deviation from unity has been seen
above $1.5T_c$ at which the spectral functions extracted by MEM also show
notable modifications.
Hence, for references, we calculate $G/G_{\text{rec}}$ at not only
temperatures where the lattice QCD results are available but also at
$T=T_c$ and at $T=1.04T_c$ where our QCD sum rule works well
and results of the lattice QCD are unambiguously anticipated.

\begin{figure}[tb]
 \includegraphics[width=3.375in]{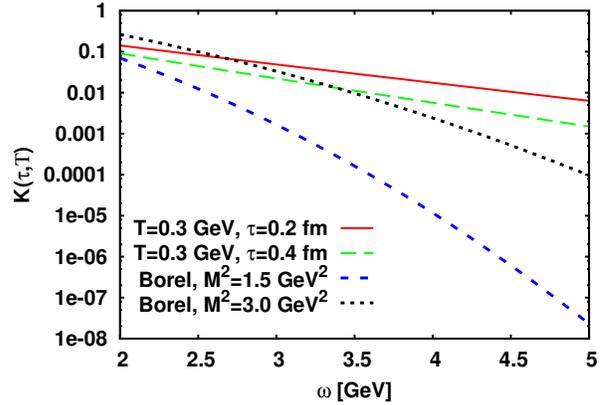}
 \caption{(color online). Integration kernels of the dispersion
 relations for the correlation functions.}
 \label{fig:kernel}
\end{figure}

Figure \ref{fig:imc_etac} displays $G/G_{\text{rec}}$ for the $P$
channel charmonium current. We plot the cases for several sets of the spectral
parameters, summarized in Appendix \ref{app:param_list}.
At $T=T_c$, $G/G_{\text{rec}}$ of $\eta_c$ can be both below and above
unity depending on the parameter set.
The parameter set of $\sqrt{s_0}=3.4$ GeV which seems most plausible
according to Fig.~\ref{fig:starkcharm} gives the closest result to unity.
We anticipate the lattice QCD gives almost unity as in the higher
temperature cases. Therefore the most plausible case gives the most
consistent result with the lattice result. Qualitatively, our results
indicate that the combination of the small decreases of both effective
threshold and mass can give a consistent result.
One sees, however, that all of the parameter sets lead to $G/G_{\text{rec}} > 1$
above $T > 1.04T_c$ in spite of the quite different spectral
parameters.
We would like to stress that $T=1.04T_c$ is the onset temperature of
$\eta_c$;
for the parameter set with the smallest $\sqrt{s_0}$ shown in the
figure, we need to introduce $\Gamma=18$ MeV to stabilize the Borel
curve. No Borel window is open for smaller $\sqrt{s_0}$.
In the two higher temperature cases, $T=1.07T_c$ and $1.09T_c$, lattice
data are shown together. Note that the lattice setup is
different between these temperatures. The latter is obtained with finer
lattice spacing. As shown in the figure, the lattice correlators do not
exhibit sizable changes from unity, that would be interpreted as no
spectral modification at these temperatures.

These disagreements with the lattice QCD results at $T \geq 1.04T_c$ do not immediately mean
serious flaws in our approach. First, the applicability of our present
approach to these temperatures are marginal and thus our results will be
quantitatively improved by including correction mentioned
above. Second, our model spectral function might be too
simple to make such a comparison. This simplification does not matter in
the QCD sum rule approach due to the strong suppression of the high energy part in the Borel transformed dispersion relation, but
might cause this defect in the correlator because of its sensitivity to
the continuum. Figure \ref{fig:kernel} displays the kernel in the
dispersion relation \eqref{eq:kernel} with typical values of parameters
together with the kernel in the Borel transformation
\eqref{eq:dispersion} where we rewrite the formula so that integration
variable is $\omega=\sqrt{s}$ and integrand has a form of
$K(\omega,M^2)\rho(\omega)$. One sees that the Borel transformation
suppresses the high energy part of the spectral function much more strongly
than the heat kernel \eqref{eq:kernel}. Note that temperature effects on
the heat kernel are small around $T_c$; it almost behaves like
$e^{-\omega\tau}$. This fact means that while the detailed
structure of the continuum and the excited states do not affect the
property of the pole part in the QCD sum rule approach, they do
contribute non-trivially to the imaginary correlator.   For example,
compared to the realistic situation,
we have neglected the contributions from the excited states ($\eta_c(2S)$ in $P$ channel) in
the model spectral function. While such approximation does not affect the
lowest pole in the QCD sum rule analysis, it will result in the smaller continuum
threshold than a realistic value to compensate for the missing state. This discrepancy between effectively small continuum
threshold and existence of excited states will result in different
imaginary time correlators.
If the $2S$ state melts just above (or below) $T_c$, our parametrization
is in fact better above $T_c$. Hence, if one takes into account the excited state
contribution explicitly,  $G_{\text{rec}}(\tau,T)$ will increase while
$G(\tau,T)$ remain unchanged, thus the ratio will now become close to
unity.
Indeed the effect of the excited state has been examined in the context
of a temperature dependent potential model in Ref.~\cite{mocsy06}.
The authors found that the $\eta_c(2S)$ state reduces $G/G_{\text{rec}}$
10--20 \% because the $2S$ state melts above $T_c$ thus increases
$G_{\text{rec}}$. One sees that the observed reduction is also relevant for
our case. Although the result of Ref.~\cite{mocsy06} shows the $2S$
states melts at $T_c$ thus our agreement at this temperature might be
changed to other set of parameters, the spectral property of the excited
states at finite temperature is not clear yet. If the $2S$ state
survives at $T_c$ and our analysis catches essential points for the
correlator, the difference between $T=T_c$ and $T > T_c$ indicates that the
$2S$ state dissolves at $T_c < T < 1.04T_c$.

\begin{figure}[t]
 \includegraphics[width=3.375in]{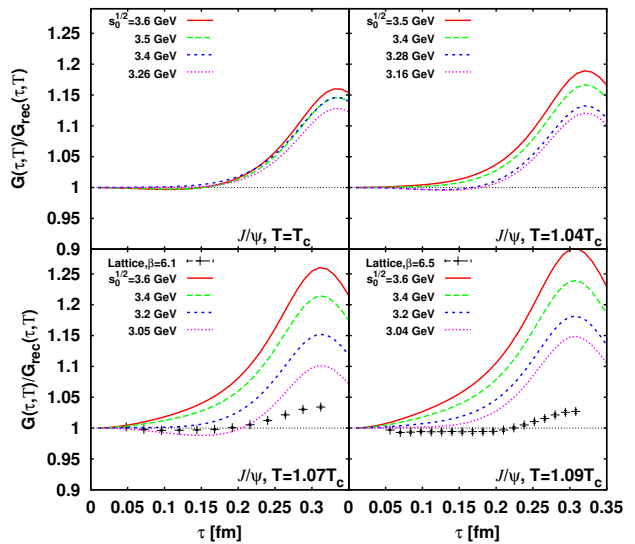}
 \caption{(color online). $G/G_{\text{rec}}$ for the $V$ channel
 charmonium current.}
 \label{fig:imc_cv}
\end{figure}

In Fig.~\ref{fig:imc_cv}, we display the results of
$G/G_{\text{rec}}$ for the vector channel ($J/\psi$).
One sees a deviation from unity starting at small $\tau$ region. This
behavior is caused by the zero mode contribution
\cite{umeda07,alberico08,mocsy08}.
At $T=1.07T_c$, our $G/G_{\text{rec}}$ with $\sqrt{s_0} \leq 3.2$ GeV
agrees the lattice results at $\tau< 0.15$ fm in which the zero mode
contribution is negligible. This means that
$G/G_{\text{rec}}\simeq 1$
is achieved by a combination of the various spectral changes as seen in the
case of $\eta_c$ at $T=T_c$.
The zero mode contribution, however, overwhelms other changes at
$T=1.07T_c$ and $T=1.09T_c$, as clearly seen from the figure.
There are following possibilities;
\begin{enumerate}
 \item High energy part of the model spectral function.

       As in the case of $\eta_c$, we have neglected $\psi'(2S)$
       contribution to the $T=0$ spectral function. Including this leads to
       larger $G_\text{rec}$ thus reducing $G/G_{\text{rec}}$. There may be also the
       possibilities that we underestimated the pole modification by
       truncation of the OPE and other approximations. As for $d=6$
       contribution in the OPE, however, it is expected to reduce the spectral
       modification \cite{Kim01}.

 \item Free charm quark approximation in the zero mode contribution.

       Indeed this might be a flaw because in more realistic situation
       quarks are interacting such that zero mode spectral function is
       smeared \cite{petreczky06}. Of course, since we are looking at the
       integrated value of the spectral function, this smearing itself
       might not change the value so much. Nevertheless, there are
       further ambiguities in the zero mode calculation such as the
       value of the charm quark mass; within the quasiparticle picture,
       the thermal effect will effectively increase the quark mass such
       that the zero mode contribution is reduced according to the thermal
       distribution.
       As we shall see below, there is clearly something that cannot be
       understood with the free charm description in the zero mode.
       To avoid these ambiguities, subtraction of the zero mode
       contribution by taking derivative the imaginary
       time correlator and then looking at the ratio \cite{umeda07,mocsy08} will
       provide useful information.
\end{enumerate}

\begin{figure}[t]
 \includegraphics[width=3.375in]{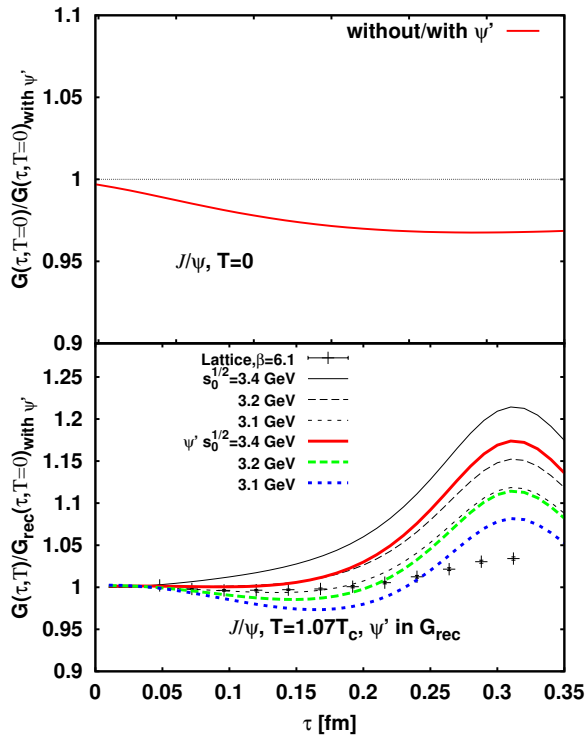}
 \caption{(color online). Effect of $2S$ state in the imaginary time
 correlator. Upper panel : ratio of the correlator $G$ without $\psi'$
 to that with $\psi'$ at $T=0$. Lower panel : $G/G_{\text{rec}}$ at
 $T=1.07T_c$ with $\psi'$ contribution being included in
 $G_{\text{rec}}$. Thin lines are the same ones shown in
 Fig.~\ref{fig:imc_cv} while thick lines denote the cases in which $\psi'$
 contribution is included in $G_{\text{rec}}$.}
 \label{fig:psiprime}
\end{figure}
In the vector channel, the first possibility can be explicitly checked
by including $\psi'$ contribution to the continuum part of the model
spectral function at $T=0$. Namely,
\begin{equation}
 \text{Im}\Pi^{\psi'}(s) = f'\delta(s-m_{\psi'}^2),
\end{equation}
with $f' = 0.276$ GeV$^2$ being obtained from the leptonic decay width given by the Particle Data Group \cite{pdg2008}, is added to the
phenomenological side \eqref{eq:phen}.
With this implementation, we found that the resultant $J/\psi$ mass
changes only 0.3\% (3.05 GeV) while the continuum threshold increases
from $\sqrt{s_0}$=3.54 GeV to 3.93 GeV. On the other hand, we also
found that incorporating the $\psi'$ to the
spectral function in the dispersion relation of the imaginary time
correlator \eqref{eq:imc} leads to a sizable change.
This fact exactly demonstrates our expectation discussed above; while
the spectral property deduced from the QCD sum rule is independent of
detailed structure of the higher energy part of the model spectral
function as long as the pole dominance is well satisfied, the imaginary
time correlator receives sizable change. Note that the physical meaning of the
threshold parameter becomes different if one takes the excited state
into account. When one includes it, now $\sqrt{s_0}$ can be regarded as
physically more relevant threshold while it represents an effective one
controlling the contribution from the high energy part other than the
lowest pole in the dispersion relations. For example,
$\sqrt{s_0}$ in Fig.~\ref{fig:starkcharm} means a merely effective
threshold parameter since we have neglected the excited state
contribution throughout the calculation. If $\psi'$ melts above $T_c$,
$\sqrt{s_0}$ can be now regarded as more physical one. In this case
the true threshold might vary more rapidly from $3.93$ GeV to a lower
value in the vicinity of $T_c$.

Figure \ref{fig:psiprime} shows the effect of $\psi'$ on the imaginary time
correlator. In the upper panel, we compare two $G(\tau,T=0)$, with and
without $\psi'$ contribution. One sees 3\% reduction of the ratio, which
means that inclusion of the $\psi'$ gives a enhancement large enough to affect
$G/G_{\text{rec}}$ comparison. The resultant $G/G_{\text{rec}}$ are
shown in the thick lines in the lower panel, where one sees the apparent
reduction of $G/G_{\text{rec}}$ when including $\psi'$. Nevertheless,
$\tau$ dependence is still governed by the zero mode contribution, on which
we will give further consideration.

\begin{figure}[t]
 \includegraphics[width=3.375in]{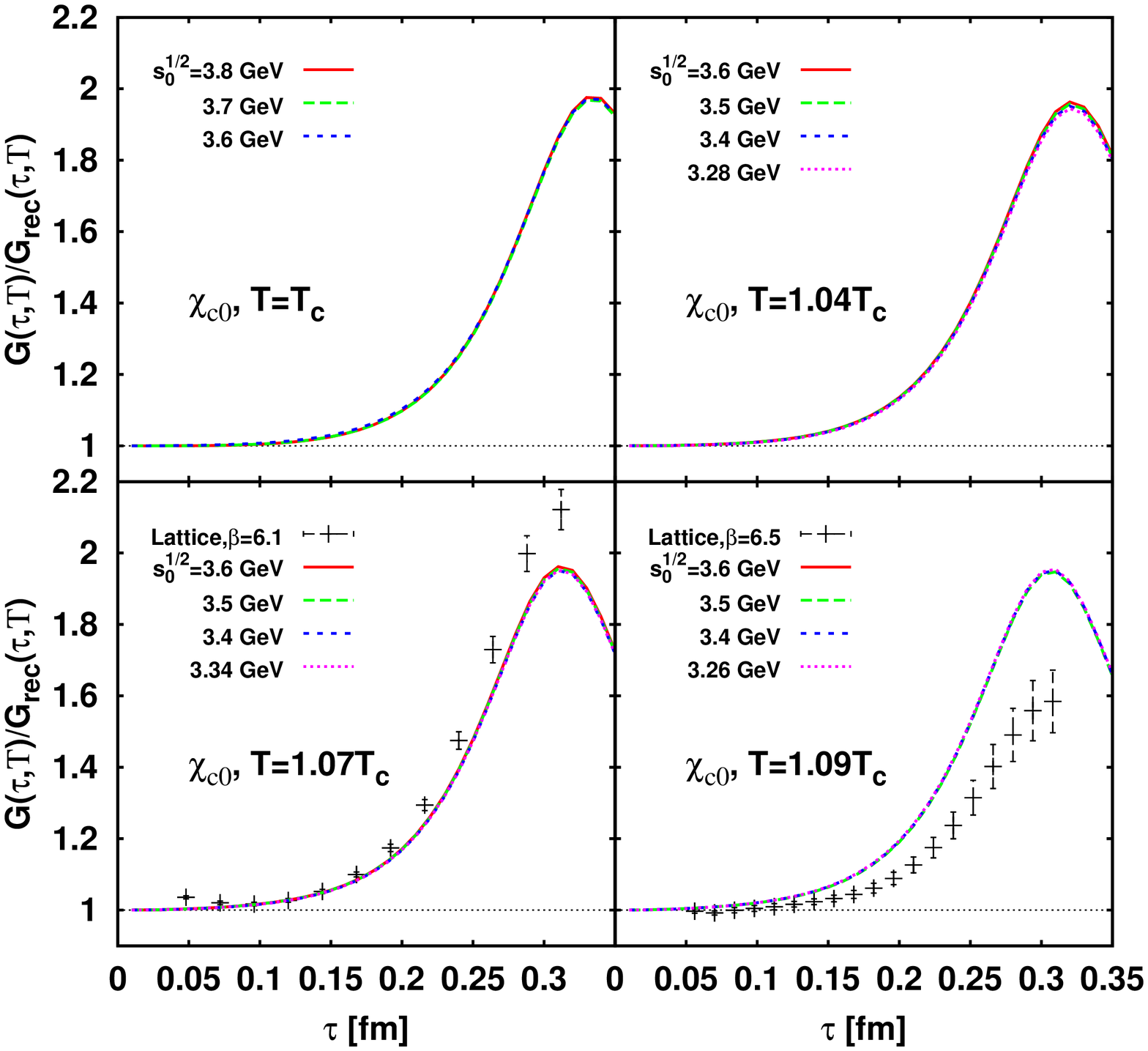}
 \caption{(color online). Same as Fig.~\ref{fig:imc_etac}, but for $S$
 channel current.}
 \label{fig:imc_chic0}

 \includegraphics[width=3.375in]{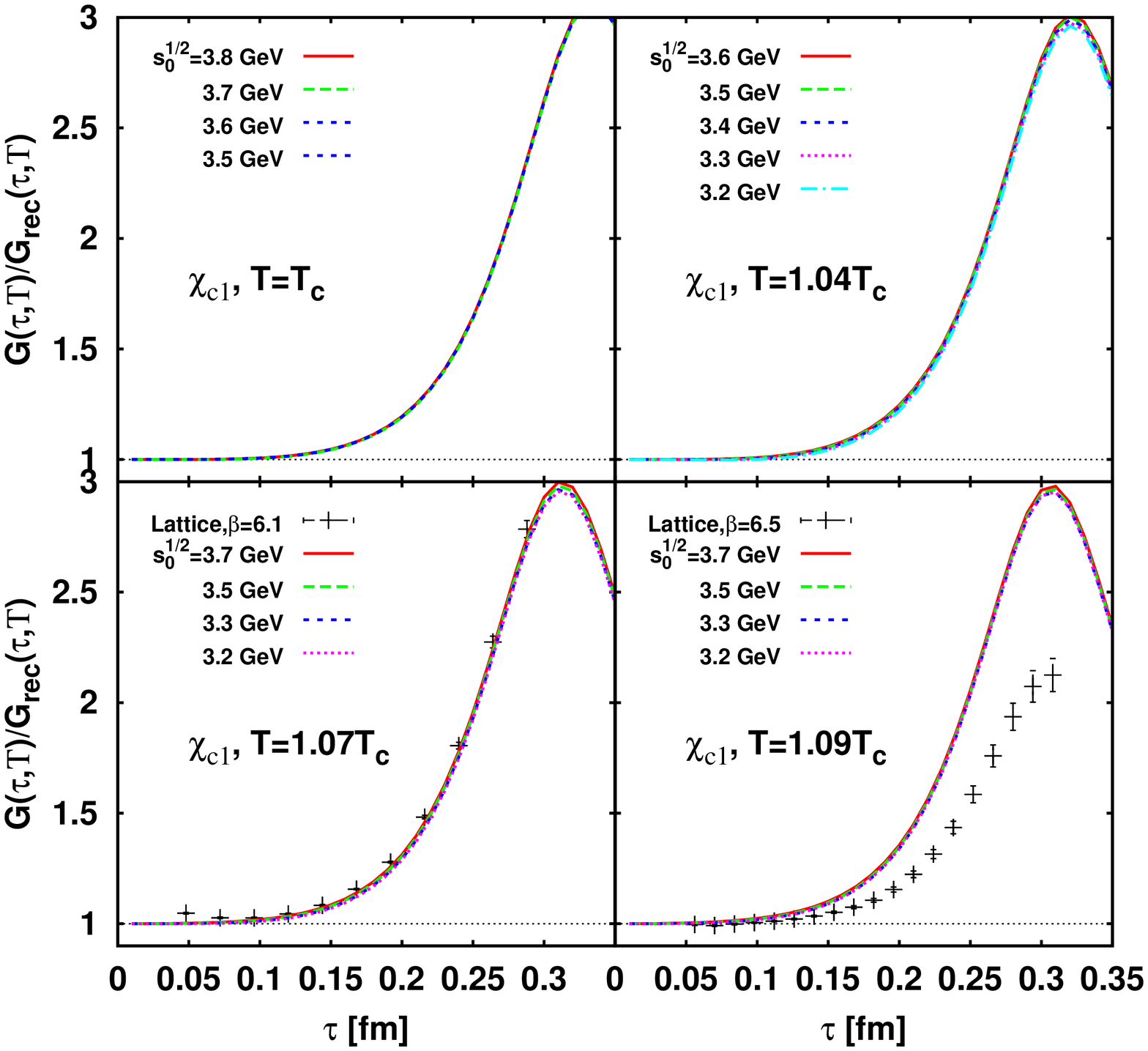}
 \caption{(color online). Same as Fig.~\ref{fig:imc_etac}, but for $A$
 channel current.}
 \label{fig:imc_chic1}
\end{figure}
\begin{figure}[t]
 \includegraphics[width=3.375in]{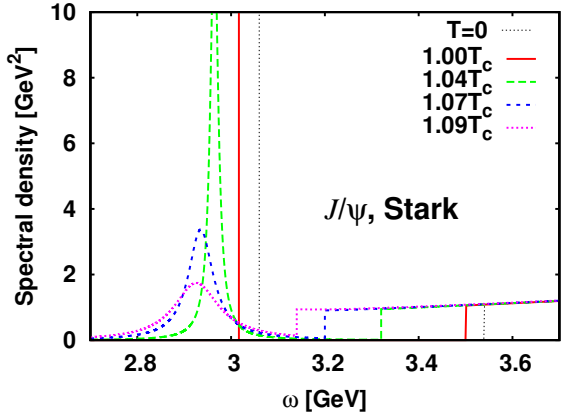}
 \caption{Spectral density of $J/\psi$ at the best fit to the lattice
 correlator.}
 \label{fig:spectral}
\end{figure}

As for the second possibility, one way to check the consistency is to
look at other channels.
We display the imaginary time correlators of the scalar and the axial-vector
channels in Figs.~\ref{fig:imc_chic0} and \ref{fig:imc_chic1},
respectively. Similarly the spectral parameters are summarized in Table
\ref{tbl:sumruleresult_chic0} and \ref{tbl:sumruleresult_chic1}.
These two channels show quite similar behavior so that
the following discussion can be applied for both cases.
First, no sizable difference among various parameter sets is seen as
indicated by the complete overlaps of the lines.
One sees the clear effect of the zero mode contribution and
its agreement with the lattice results at $T=1.07T_c$ contrary to the
$V$ channel case. This might be partly attributed to the absence of the
$2S$ state contribution below the continuum threshold in the $S$ and the $A$
channels; \textit{i.e.}, the single pole plus continuum ansatz at $T=0$
is a better approximation in these channels than in the $V$ and the $P$ channels.
At $T=1.09T_c$, however, this agreement is lost though qualitatively the lattice results
indicate the dominance of the zero mode. One sees the value of
$G/G_{\text{rec}}$ is smaller at $T=1.09T_c$ than at $T=1.07T_c$ in the
lattice results. If the spectral modification of the pole and the
continuum part does not differ so much between these temperatures,
this result seems to indicate the \textit{smaller} zero mode
contribution at higher temperature. This cannot be understood within the
free charm quark approximation in which zero mode contribution increases
as temperature increases if the charm quark mass is constant. Therefore,
although our results
show agreement with the lattice results in the $\tau$ range where zero
mode contribution is relatively small, we cannot draw definite
conclusion on the quantitative correctness of the zero mode from the
results.
\begin{figure}[t]
 \includegraphics[width=3.375in]{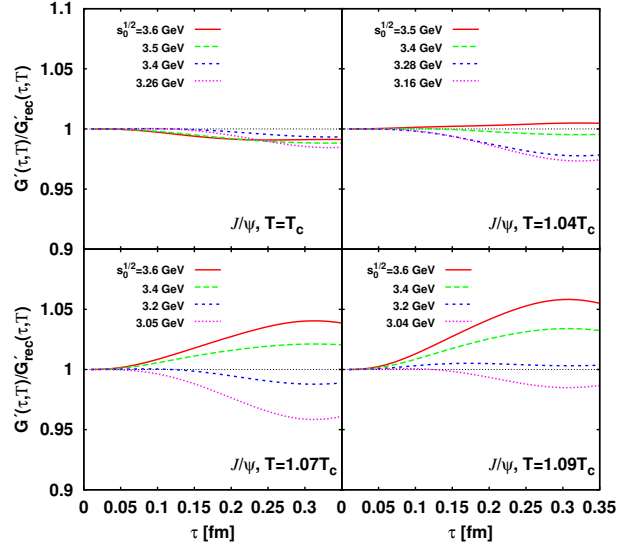}
 \caption{(color online). Ratio of the derivative of the imaginary time correlator
 $G'/G'_{\text{rec}}$ for the $V$ channel. Shown parameter sets are the
 same as in Fig.~\ref{fig:imc_cv}.}
 \label{fig:imcd_CV}
\end{figure}
\begin{figure}[t]
 \includegraphics[width=3.375in]{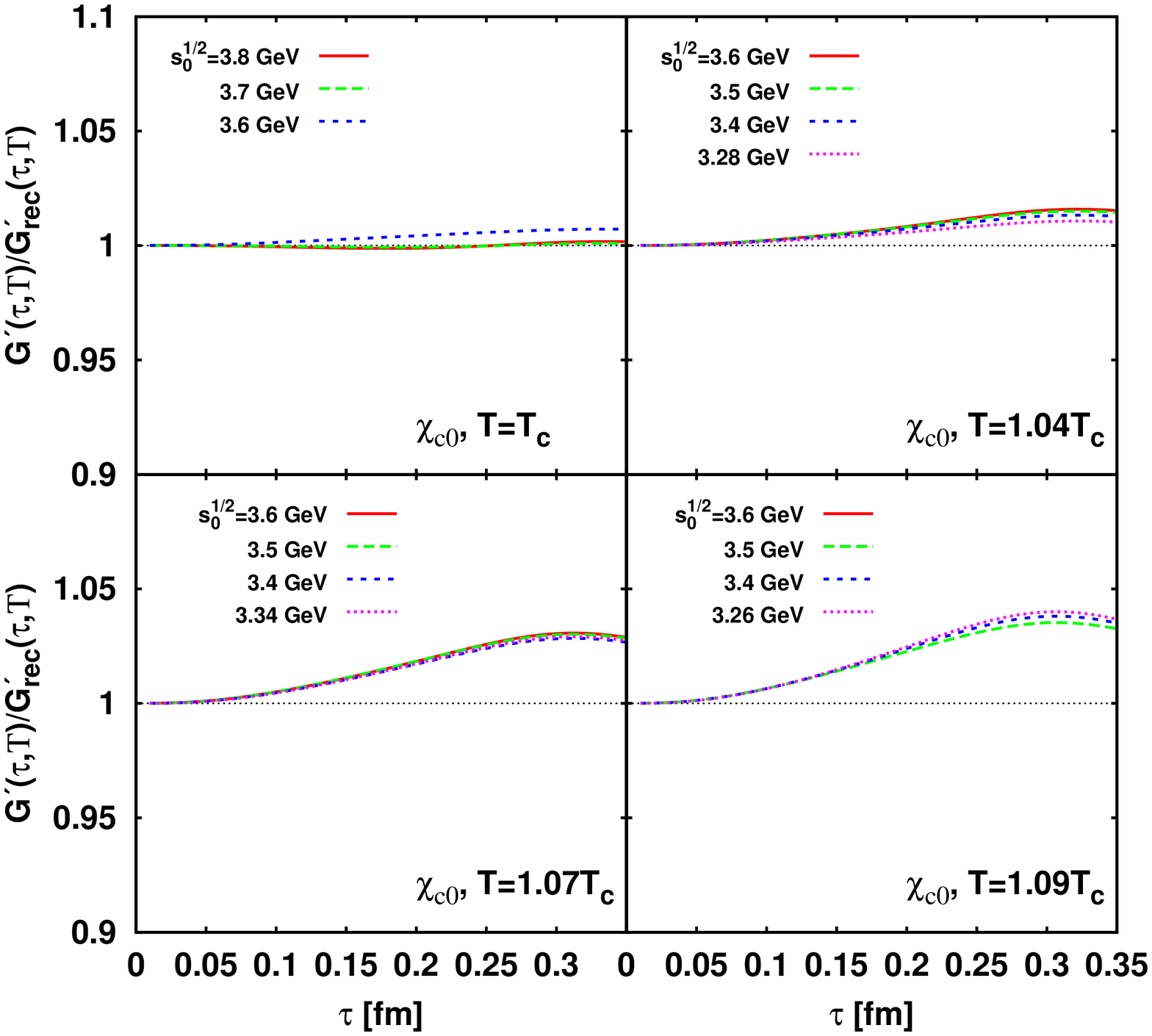}
 \caption{(color online). Same as Fig.~\ref{fig:imcd_CV}, but for $S$
 channel.}
\end{figure}
\begin{figure}[t]
 \includegraphics[width=3.375in]{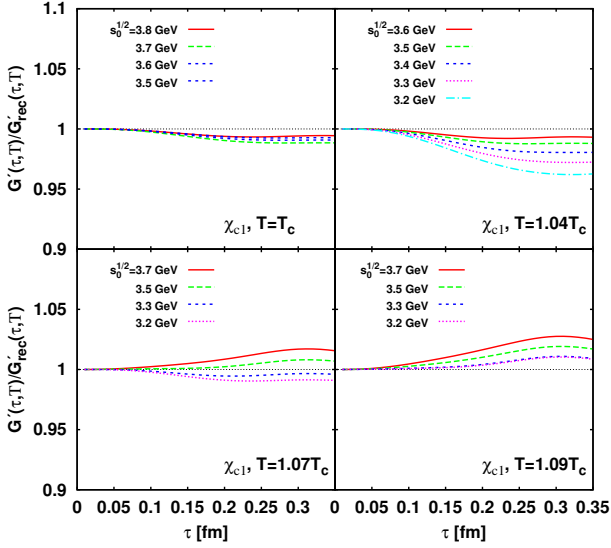}
 \caption{(color online). Same as Fig.~\ref{fig:imcd_CV}, but for $A$
 channel.}
 \label{fig:imcd_CA}
\end{figure}
One way to avoid the difficulty of the zero mode is to evaluate the
derivative of the imaginary time correlator with respect to $\tau$ \cite{umeda07}.
Figures \ref{fig:imcd_CV}--\ref{fig:imcd_CA} show the ratio of the
derivative of the imaginary time correlator, $G'/G'_{\text{rec}}$, in
which the zero mode contribution is absent.
One sees different tendency from $G/G_{\text{rec}}$ such that
$G'/G'_{\text{rec}}\simeq 1$ within
3\% for certain sets of the spectral parameters at
$T < T_{\text{onset}}$. This strongly supports robustness of our results
at these temperatures since the lattice computations, although not
available at these temperature, are expected to be unity also. We would
like to note that the parameter set close to the ones constrained from
the Stark effect gives $G'/G'_{\text{rec}}$ closest to unity in the case
of $J/\psi$.  For illustrative purpose, we display the spectral
density of $J/\psi$ constructed from Eq.~\eqref{eq:imc_rho_pc} in Fig.~\ref{fig:spectral} with the same parameter
sets as those in Fig.~\ref{fig:starkcharm}.

As demonstrated in Ref.~\cite{mocsy08}, effect of the threshold
enhancement \cite{cabrera07:_t,laine07:_qcd} seems important for
understanding the relation between
the lattice measurement which gives $G/G_{\text{rec}}\simeq 1 $ and
model calculations such as potential models. In $\eta_c$ and $J/\psi$,
it explains successfully the lattice data at $T=1.2T_c$.
Note that this is beyond our regime in the present work;
we cannot apply our method at this temperature since we could not have
the reliable Borel stability beyond $T_{\text{onset}}$.
We can still see some indication of such an effect in the present data.
One notes that the Borel curve becomes flatter and flatter, according to
$\sqrt{\chi^2}$, as $\sqrt{s_0}$ decreases above $T=T_{\text{onset}}$.
(See Tables \ref{tbl:sumruleresult_jpsi} and
\ref{tbl:sumruleresult_etac}).
It has minimum at the smallest $\sqrt{s_0}$ at which the Borel window is
about to close. This fact means that if one relaxes the criterion for
the pole dominance, one still obtains the stability for smaller $\sqrt{s_0}$.
Though any extrapolation to higher temperature cannot be
reliable due to the missing effects as discussed, one might expect
the threshold to be close to the mass thus the spectral function
exhibits the threshold
enhancement. Whereas direct confirmation will not be possible with the present
method by construction since pole dominance will be badly violated,
this behavior of $\sqrt{\chi^2}$ may indicate a connection to the
higher temperature regime which has been investigated only through the
potential models so far.

\subsubsection{Bottomonium}

\begin{figure}[t]
 \includegraphics[width=3.375in]{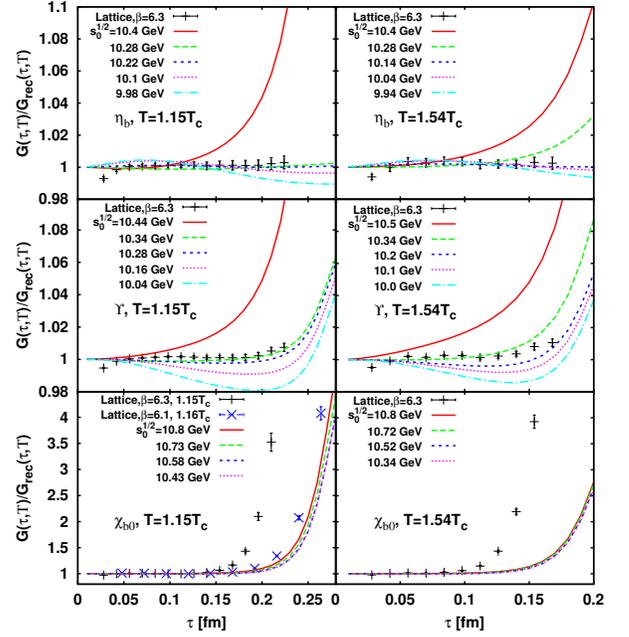}
 \caption{(color online). $G/G_{\text{rec}}$ for bottomonium currents.
 Lattice results taken from Ref.~\cite{jakovac07}. Left and right columns
 stand for $T=1.15T_c$ and $T=1.54T_c$ cases while each
 rows denote $P$, $V$, and $S$ channels from top to bottom, respectively.}
 \label{fig:imc_B}
\end{figure}

We compute $G/G_{\text{rec}}$ for the bottomonium currents in the same
manner. Figure \ref{fig:imc_B} displays the results
at $T=1.15T_c$ and $T=1.54T_c$ of which lattice results are available
with the finest lattice spacing in Ref.~\cite{jakovac07}. As in the charmonium cases, the lattice results show little deviation from
unity.
In the $S$-waves, while our results show variations among the parameter sets
there is a  certain range of the effective threshold parameter of which
parameter set gives $G/G_{\text{rec}}\simeq 1$. The best agreement seems to lie between
the data set with vacuum $\sqrt{s_0}$ (green, long-dashed) and that with
Stark effect results (blue, short-dashed). The difference of these two
cases is tiny at $T=1.15T_c$ but sizable at $1.54T_c$, indicating
the possibility of discriminating the spectral change from the imaginary
time correlator. As before, one has to consider excited states
to give a definite conclusion. The interpretation of the behavior of
$G/G_{\text{rec}}$ depends on
whether such states below threshold survive or not at these
temperatures. From these agreements, a possible interpretation is that
excited state are still surviving even at $T=1.54T_c$ since a potential
model calculation shows larger reduction of $G/G_{\text{rec}}$ by
including $2S$ and $3S$ states in $\eta_b$ than the variation seen in
the figure \cite{mocsy06}.
The result of $\chi_{b0}$ state is again dominated by the zero
mode contribution as was the case for the $\chi_{c0}$. One sees that even lattice
results show a difference between that at $T=1.15T_c$ with $\beta=6.1$
and that at $T=1.16T_c$ with $\beta=6.3$, indicating the difficulty in
quantifying the calculations. Therefore we compute $G'/G'_{\text{rec}}$
as was done in the charmonium cases.
\begin{figure}[t]
 \includegraphics[width=3.375in]{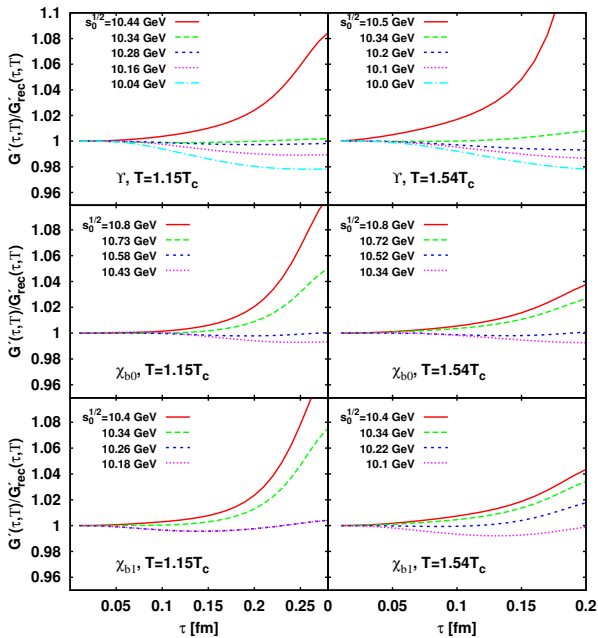}
 \caption{(color online). $G'/G'_{\text{rec}}$ for bottomonium currents.
 Left and right columns stand for $T=1.15T_c$ and $T=1.54T_c$ cases while each
 rows denote $V$, $S$, and $A$ channels from top to bottom, respectively.}
 \label{fig:imcd_B}
\end{figure}
Figure \ref{fig:imcd_B} shows the results for $G'/G'_{\text{rec}}$ of
$V$, $S$, and $A$ channels.  As before, one sees that
$G'/G'_{\text{rec}}$ is sensitive to the variation of the spectral
parameters and there exist certain ranges which give
$G'/G'_{\text{rec}}\simeq 1$. Precise determination of this
quantity will be useful for constraining the spectral changes.
One sees that dropping of both the mass and the
continuum threshold is consistent with $G'/G'_{\text{rec}}\simeq 1$
and that the result from the second order Stark effect again fits well in the case
of $\Upsilon$.

\section{Summary and outlook}
\label{sec:summary}

We have analyzed the spectral changes of heavy quarkonia in the hot
environment in a systematic way based on the QCD sum rule with Borel
transformation technique. We have taken into account possible changes of
the continuum spectrum by a temperature dependent effective continuum
threshold. Although the temperature dependence of the OPE side
\eqref{eq:borel_moment} allows various combination of the changes of the
spectral parameters, we have given the constraints among them by an
optimization procedure which has been widely used in QCD sum rule
applications. We found that instability of the Borel curve in the
$\Gamma=0$ limit caused by the change of the gluon condensates can be
cured by introducing a width although neglected contributions from
higher dimensional operators might be non-negligible at the small Borel mass.
The results,
Figs.\ref{fig:qcdsr_etac}--\ref{fig:qcdsr_chic1} and
\ref{fig:qcdsr_etab}--\ref{fig:qcdsr_chib1}, show the behaviors of the
spectral parameters with respect to the change of temperature through
the gluon condensates. As already argued in previous literatures
\cite{morita_jpsifull,song09,lee_morita_stark}, charmonia exhibit the
critical behavior in at least one of the spectral parameters. Note that
although the effective continuum threshold shares the effect from the
change of the gluon condensates and thus reduces the change of other
parameters, it is basically linked with the mass as seen in the
Figs.~\ref{fig:qcdsr_etac}--\ref{fig:qcdsr_chic1} and
\ref{fig:qcdsr_etab}--\ref{fig:qcdsr_chib1} as a result of stabilizing
the Borel curve. While we do not taken into
account the excited state explicitly, property of the lowest pole is
not affected by this simplification since it is imposed on
the effective threshold parameter. When one of the spectral
parameters remains unchanged, rapid change of other parameters is
inevitable. We found there is an onset temperature for each channel
at which broadening must occur. Although the value of this temperature
is affected by the assumption on the Borel window and does not
necessarily mean there is no significant broadening below, the combined
analysis with the QCD second order Stark effect shows that $\eta_c$ and
$J/\psi$ do not likely have significant width below $T_c$ and
all the spectral parameters of $J/\psi$ and $\eta_c$ change abruptly in the
vicinity of $T_c$.
The same analysis procedure for the bottomonia shows the little change
around $T_c$ because of the much heavier quark mass, but eventually
shows sizable changes with increasing temperature.

Such changes should be obtained from the lattice QCD
also, while spectral analyses based on MEM do not have sufficient
resolution. Therefore, we have computed the imaginary time correlator
which is the basis of the MEM analyses
and its derivative with respect to the Euclidean time $\tau$ by putting
the phenomenological side of the QCD sum rule analyses as a simple model
of the spectral function. Then, we
take the ratio $G/G_{\text{rec}}$ in order to see the temperature effect
on the spectral function, as done in the lattice analyses and potential
model calculations. We have demonstrated that the results obtained in the
lattice calculation, namely $G/G_{\text{rec}}\simeq 1$, do not always mean
 absence of the spectral changes, but can mean  a mixture of some sets of the
change of the spectral properties. Similar observation has
been done in some potential model analyses \cite{cabrera07:_t,mocsy08}
but they are at higher temperatures and mainly focused on the threshold
enhancement. We showed that the rather small modification compared to the
bound state masses gives the sizable change in $G/G_{\text{rec}}$ and the
constrained parameter sets by the QCD sum rule can lead to
$G/G_{\text{rec}}\simeq 1$.
We have also pointed out that while similarity in the dispersion
relation exists between the Borel-transformed current correlation function
\eqref{eq:dispersion} and the imaginary time correlator \eqref{eq:imc},
the former is much more dominated by the pole contribution. As a result,
the QCD sum rule analysis is not much affected by taking into account the known  excited states
explicitly, while the imaginary time correlator shows a small but
significant change over the uncertainties of the lattice QCD results.
Due to this property and the poorly known spectral function near
$2m_c$ threshold region in the case of finite $\Gamma$, more work is needed
to give a precise quantitative determination of the spectral
parameters.

Before closing, we comment on implication for the full QCD case
by repeating our argument in
Refs.~\cite{morita_jpsifull,song09,lee_morita_stark} since the present
work is based on quenched QCD.
If one takes into account light dynamical quarks in the OPE, they appear
as light quark condensate contributions and change of the temperature
dependence of the gluon condensates in the OPE. The former can be safely
neglected since it is at order $\alpha_s^2(q^2)$. The latter has been
shown to lead smoother temperature dependence of $G_0$ near $T_c$
\cite{morita_jpsifull} reflecting the crossover nature of the
transition. One also expects the similar change in $G_2$.
Since the OPE side depends on temperature through these condensates,
this will result in smoother spectral changes than those shown in the present
work.  While the spectral changes become smoother, the actual magnitude
of changes at $T_c$ might not differ so much since the reduction of the
scalar condensate at $T=T_c$ is almost the same as that for the quenched
case \cite{morita_jpsifull}. This also implies that the broadening
below $T_c$ could be negligible.
On the other hand, the moderate decrease of the condensates above
$T_c$ may lead to higher onset temperatures in the full QCD case.

The results shown above strongly indicate that our main results, the mass
shifts and width broadening induced by the QCD phase transition, might be
also realized in lattice QCD simulation. The agreement between the sum
rule constrants and the lattice correlator ratio is obtained on the basis
of the effect of the known excited state on the lattice correlator and
the zero mode contribution of the free heavy quarks. The best set of the parameters
depends on such external assumptions.
Further assessments of these quantities as well as continuum spectrum at finite
temperature \cite{burnier09:_heavy} will be required to check the
consistency more quantitatively.

\begin{acknowledgments}
 The authors would like to thank P.~Petrezcky and T.~Hatsuda for valuable discussion and
 suggestions. K.M. would like to acknowledge A.~Velytsky for providing
 him numerical tables of lattice results of $G/G_{\text{rec}}$. We thank
 Institute for Nuclear Theory at the University of Washington for its
 hospitality and the Department of Energy for partial support during the
 ``Joint CATHIE-INT mini-program Quarkonium in Hot Media : from QCD to
 Experiment INT-09-42W'' where a part of this work was completed.
 K.M. is indebted to S.~Muroya and N.~Suzuki for fruitful discussion and
 their kind hospitality during his visit to Matsumoto University.
 This work was supported by the Korean Ministry of Education through the
 BK21 Program and KRF-2006-C00011.
\end{acknowledgments}

\appendix

\section{Borel transformed Wilson coefficients}
\label{app:wilson}

In this appendix, we list the Wilson coefficients seen in
Eq.~\eqref{eq:borel_moment} originally obtained by Bertlmann in
Ref.~\cite{bertlmann82}.

Hereafter
\begin{align}
 c_2&= \frac{\pi}{2}-\frac{3}{4\pi}\\
 c_1&= \frac{\pi}{3}+\frac{1}{2} c_2\\
 c_3&= \frac{\pi}{2}-\frac{3}{\pi}
\end{align}
and $G(a,b,\nu)$ is the Whittaker function defined as
\begin{equation}
 G(a,b,\nu) \equiv \frac{1}{\Gamma(b)}\int_{0}^{\infty}ds \, e^{-s}s^{b-1}(\nu+s)^{-a},\label{eq:whittaker}
\end{equation}
with $\Gamma(b)$ being the Gamma function.

\begin{widetext}
\subsection{V channel}
\begin{align}
 A(\nu)&=
 \frac{3}{16\pi^{3/2}}\frac{4m_h^2}{\nu}\Whittaker{\frac{1}{2}}{\frac{5}{2}}{\nu},\\
 a(\nu)&=\frac{4}{3\sqrt{\pi}
 \Whittaker{\frac{1}{2}}{\frac{5}{2}}{\nu}}
 \left[ \pi-c_1 \Whittaker{1}{2}{\nu}+\frac{1}{3}c_2 \Whittaker{2}{3}{\nu} \right]
 -c_2-\frac{4\ln 2}{\pi}h(\nu),\\
 h(\nu)&= \nu
 \frac{\Whittaker{\frac{1}{2}}{\frac{3}{2}}{\nu}}{\Whittaker{\frac{1}{2}}{\frac{5}{2}}{\nu}},\\
 b(\nu)&=
 -\frac{\nu^2}{2}\frac{\Whittaker{-\frac{1}{2}}{\frac{3}{2}}{\nu}}{\Whittaker{\frac{1}{2}}{\frac{5}{2}}{\nu}},\\
 c(\nu)&=
 b(\nu)-\frac{2}{3}\nu^2\frac{G(\frac{3}{2},\frac{3}{2},\nu)}{G(\frac{1}{2},\frac{5}{2},\nu)}\\
\end{align}
Derivatives with respect to $\nu$ are used in Eq.~\eqref{eq:sumrule}.
\begin{align}
 A'(\nu)&= -\frac{3m_h^2}{4\pi^{3/2}\nu}\left[
 \frac{\Whittaker{\frac{3}{2}}{\frac{5}{2}}{\nu}}{2}+\frac{1}{\nu}\Whittaker{\frac{1}{2}}{\frac{5}{2}}{\nu}\right],\\
 a'(\nu)&= -\frac{4}{3\sqrt{\pi}\Whittaker{\frac{1}{2}}{\frac{5}{2}}{\nu}}\left\{ -c_1
 \Whittaker{2}{2}{\nu}+ \frac{2}{3}c_2 \Whittaker{3}{3}{\nu}
 -\frac{\Whittaker{\frac{3}{2}}{\frac{5}{2}}{\nu}}{2\Whittaker{\frac{1}{2}}{\frac{5}{2}}{\nu}}
 \left[ \pi-c_1
 \Whittaker{1}{2}{\nu} + \frac{1}{3}c_2 \Whittaker{2}{3}{\nu}
 \right] \right\} \nonumber\\
 & - \frac{4\ln 2}{\pi}h'(\nu),\\
 h'(\nu)&=
 \frac{\Whittaker{\frac{1}{2}}{\frac{3}{2}}{\nu}}{\Whittaker{\frac{1}{2}}{\frac{5}{2}}{\nu}}-\frac{\nu}{2\left[
 \Whittaker{\frac{1}{2}}{\frac{5}{2}}{\nu} \right]^2}\left[
 \Whittaker{\frac{3}{2}}{\frac{3}{2}}{\nu}\Whittaker{\frac{1}{2}}{\frac{5}{2}}{\nu}
 -\Whittaker{\frac{1}{2}}{\frac{3}{2}}{\nu}\Whittaker{\frac{3}{2}}{\frac{5}{2}}{\nu}
 \right],\\
 b'(\nu)&= \frac{-\nu}{\Whittaker{\frac{1}{2}}{\frac{5}{2}}{\nu}}\left[
 \Whittaker{-\frac{1}{2}}{\frac{3}{2}}{\nu}+\frac{\nu\Whittaker{\frac{1}{2}}{\frac{3}{2}}{\nu}}{4}
 + \frac{\nu
 \Whittaker{-\frac{1}{2}}{\frac{3}{2}}{\nu}\Whittaker{\frac{3}{2}}{\frac{5}{2}}{\nu}}{4\Whittaker{\frac{1}{2}}{\frac{5}{2}}{\nu}}
 \right],\\
 c'(\nu)&= b'(\nu)-\frac{4}{3}\nu
 \frac{\Whittaker{\frac{3}{2}}{\frac{3}{2}}{\nu}}{\Whittaker{\frac{1}{2}}{\frac{5}{2}}{\nu}} -
 \frac{2\nu^2}{3
 \left[\Whittaker{\frac{1}{2}}{\frac{5}{2}}{\nu}\right]^2}\left[\frac{1}{2}\Whittaker{\frac{3}{2}}{\frac{3}{2}}{\nu}
 \Whittaker{\frac{3}{2}}{\frac{5}{2}}{\nu}-\frac{3}{2}\Whittaker{\frac{5}{2}}{\frac{3}{2}}{\nu}\Whittaker{\frac{1}{2}}{\frac{5}{2}}{\nu}\right],
\end{align}
where we used
\begin{equation}
 \frac{\partial}{\partial \nu}G(b,c,\nu) = -bG(b+1,c,\nu).
\end{equation}

\subsection{P channel}

\begin{align}
 A(\nu)&=
 \frac{3}{16\pi^{3/2}}\frac{4m_h^2}{\nu}\Whittaker{\frac{1}{2}}{\frac{3}{2}}{\nu},\\
 a(\nu)&=\frac{4}{3\sqrt{\pi}\Whittaker{\frac{1}{2}}{\frac{3}{2}}{\nu}}
 \left[ \pi -\frac{1}{2}c_1 \Whittaker{1}{2}{\nu} \right] -c_2
 +\frac{1}{\pi}\left[\frac{8}{3}+S(\nu)\right]-\frac{4 \ln
 2}{\pi}h(\nu),\\
 h(\nu)&=
 \nu\frac{\Whittaker{\frac{1}{2}}{\frac{1}{2}}{\nu}}{\Whittaker{\frac{1}{2}}{\frac{3}{2}}{\nu}},\\
 S(\nu)&=
 -\frac{4\Whittaker{\frac{3}{2}}{\frac{3}{2}}{\nu}+\frac{5}{6}\Whittaker{\frac{3}{2}}{\frac{5}{2}}{\nu}}{\Whittaker{\frac{1}{2}}{\frac{3}{2}}{\nu}},\\
 b(\nu)&= -\frac{1}{2}\nu
 \frac{\Whittaker{-\frac{3}{2}}{\frac{3}{2}}{\nu}-6\Whittaker{-\frac{1}{2}}{\frac{3}{2}}{\nu}}{\Whittaker{\frac{1}{2}}{\frac{3}{2}}{\nu}}\label{eq:b_ps}\\
 c(\nu)&=
 b(\nu)-4\nu\frac{\Whittaker{-\frac{1}{2}}{\frac{1}{2}}{\nu}}{\Whittaker{\frac{1}{2}}{\frac{3}{2}}{\nu}},\\
 A'(\nu)&= -\frac{3m_h^2}{4\pi^{3/2}\nu}\left[
 \frac{1}{2}\Whittaker{\frac{3}{2}}{\frac{3}{2}}{\nu}+\frac{1}{\nu}\Whittaker{\frac{1}{2}}{\frac{3}{2}}{\nu}
 \right],\\
 a'(\nu)&=
 \frac{2}{3\sqrt{\pi}\Whittaker{\frac{1}{2}}{\frac{3}{2}}{\nu}}\left\{
 c_2
 \Whittaker{2}{2}{\nu}+\frac{\Whittaker{\frac{3}{2}}{\frac{3}{2}}{\nu}}{\Whittaker{\frac{1}{2}}{\frac{3}{2}}{\nu}}
 \left[\pi -\frac{1}{2}c_2 \Whittaker{1}{2}{\nu}
 \right]\right\}+\frac{1}{\pi}S'(\nu)-\frac{4 \ln 2}{\pi}h'(\nu),\\
 h'(\nu)&=
 \frac{\Whittaker{\frac{1}{2}}{\frac{1}{2}}{\nu}}{\Whittaker{\frac{1}{2}}{\frac{3}{2}}{\nu}}
 -\frac{\nu}{2[\Whittaker{\frac{1}{2}}{\frac{3}{2}}{\nu}]^2}
 \left[
 \Whittaker{\frac{3}{2}}{\frac{1}{2}}{\nu}\Whittaker{\frac{1}{2}}{\frac{3}{2}}{\nu}-\Whittaker{\frac{1}{2}}{\frac{1}{2}}{\nu}\Whittaker{\frac{3}{2}}{\frac{3}{2}}{\nu}\right],\\
 S'(\nu)&=
 \frac{1}{\Whittaker{\frac{1}{2}}{\frac{3}{2}}{\nu}}\left\{
 6\Whittaker{\frac{5}{2}}{\frac{3}{2}}{\nu}+\frac{15}{12}\Whittaker{\frac{5}{2}}{\frac{5}{2}}{\nu}
 -\frac{1}{2}\frac{\Whittaker{\frac{3}{2}}{\frac{3}{2}}{\nu}}{\Whittaker{\frac{1}{2}}{\frac{3}{2}}{\nu}}
 \left[ 4\Whittaker{\frac{3}{2}}{\frac{3}{2}}{\nu}+\frac{5}{6}
 \Whittaker{\frac{3}{2}}{\frac{5}{2}}{\nu} \right] \right\},\\
 b'(\nu)&=
 \frac{-1}{2\Whittaker{\frac{1}{2}}{\frac{3}{2}}{\nu}}\left\{
 \frac{3\nu}{2}\Whittaker{-\frac{1}{2}}{\frac{3}{2}}{\nu}-3\nu
 \Whittaker{\frac{1}{2}}{\frac{3}{2}}{\nu} + \left[
 1+\frac{\nu}{2}
 \frac{\Whittaker{\frac{3}{2}}{\frac{3}{2}}{\nu}}{\Whittaker{\frac{1}{2}}{\frac{3}{2}}{\nu}}\right]
 \left[
 \Whittaker{-\frac{3}{2}}{\frac{3}{2}}{\nu}-6\Whittaker{-\frac{1}{2}}{\frac{3}{2}}{\nu}\right]\right\},\\
 c'(\nu)&=
 b'(\nu)-\frac{4\Whittaker{-\frac{1}{2}}{\frac{1}{2}}{\nu}}{\Whittaker{\frac{1}{2}}{\frac{3}{2}}{\nu}}-
 \frac{2\nu}{\left[ \Whittaker{\frac{1}{2}}{\frac{3}{2}}{\nu}
 \right]
^2} \left[
 \Whittaker{\frac{1}{2}}{\frac{1}{2}}{\nu}\Whittaker{\frac{1}{2}}{\frac{3}{2}}{\nu}
 +
 \Whittaker{-\frac{1}{2}}{\frac{1}{2}}{\nu}\Whittaker{\frac{3}{2}}{\frac{3}{2}}{\nu} \right].
\end{align}

\subsection{S channel}

\begin{align}
 A(\nu)&=
 \frac{9}{32\pi^{3/2}}\frac{4m_h^2}{\nu}\Whittaker{\frac{3}{2}}{\frac{5}{2}}{\nu},\\
 a(\nu)&= \tilde{a}(\nu) +\frac{1}{\pi}\left[
 \frac{16}{3}+S_s(\nu) \right],\\
 \tilde{a}(\nu)&=
 \frac{8}{9\sqrt{\pi}}\frac{\Whittaker{1}{2}{\nu}}{\Whittaker{\frac{3}{2}}{\frac{5}{2}}{\nu}}
 \left[ \pi -2c_3
 \frac{\Whittaker{2}{3}{\nu}}{\Whittaker{1}{2}{\nu}}\right] -\frac{2}{3}c_3-\frac{4\ln 2}{\pi}h(\nu), \\
 h(\nu)&=
 \nu\frac{\Whittaker{\frac{3}{2}}{\frac{3}{2}}{\nu}}{\Whittaker{\frac{3}{2}}{\frac{5}{2}}{\nu}},\\
 S_s(\nu)&= -\frac{4}{3\Whittaker{\frac{3}{2}}{\frac{5}{2}}{\nu}}
 \left[
 5\Whittaker{\frac{5}{2}}{\frac{5}{2}}{\nu}+\frac{3}{2}\Whittaker{\frac{7}{2}}{\frac{9}{2}}{\nu}+\left(
 5-\frac{1}{\nu} \right) \Whittaker{\frac{5}{2}}{\frac{7}{2}}{\nu}
 \right],\\
 b(\nu)&=
 -\frac{3\nu}{2\Whittaker{\frac{3}{2}}{\frac{5}{2}}{\nu}} \left[
 \Whittaker{-\frac{1}{2}}{\frac{5}{2}}{\nu}-\frac{2}{3}\Whittaker{\frac{1}{2}}{\frac{5}{2}}{\nu}
 \right],\\
 c(\nu)&=
 b(\nu)+\frac{4\nu}{3}\frac{\Whittaker{\frac{1}{2}}{\frac{3}{2}}{\nu}}{\Whittaker{\frac{3}{2}}{\frac{5}{2}}{\nu}},\\
 A'(\nu)&= -\frac{9m_h^2}{8\pi^{3/2}\nu}\left[
 \frac{3}{2}\Whittaker{\frac{5}{2}}{\frac{5}{2}}{\nu} +
 \frac{1}{\nu} \Whittaker{\frac{3}{2}}{\frac{5}{2}}{\nu} \right],\\
 a'(\nu)&= \tilde{a}'(\nu)+\frac{1}{\pi} S_s'(\nu),\\
 \tilde{a}'(\nu)&=
 -\frac{8}{9\sqrt{\pi}\Whittaker{\frac{3}{2}}{\frac{5}{2}}{\nu}}
 \left\{ -2c_3 \left[ 2\Whittaker{3}{3}{\nu} -
 \frac{\Whittaker{2}{2}{\nu}\Whittaker{2}{3}{\nu}}{\Whittaker{1}{2}{\nu}}\right]
 + \left[ \pi - 2c_3
 \frac{\Whittaker{2}{3}{\nu}}{\Whittaker{1}{2}{\nu}}\right]\right.
 \nonumber\\
 & \left.\times\left[
 \Whittaker{2}{2}{\nu}-\frac{3}{2}\frac{\Whittaker{1}{2}{\nu}\Whittaker{\frac{5}{2}}{\frac{5}{2}}{\nu}}{\Whittaker{\frac{3}{2}}{\frac{5}{2}}{\nu}}\right]\right\}-\frac{4\ln
 2}{\pi}h'(\nu),\\
h'(\nu)&=
 \frac{\Whittaker{\frac{3}{2}}{\frac{3}{2}}{\nu}}{\Whittaker{\frac{3}{2}}{\frac{5}{2}}{\nu}}
 -\frac{\nu}{2[\Whittaker{\frac{3}{2}}{\frac{5}{2}}{\nu}]^2}
 \left[
 \Whittaker{\frac{5}{2}}{\frac{3}{2}}{\nu}\Whittaker{\frac{3}{2}}{\frac{5}{2}}{\nu}
 -\Whittaker{\frac{3}{2}}{\frac{3}{2}}{\nu}\Whittaker{\frac{5}{2}}{\frac{5}{2}}{\nu}\right],\\
 S_s'(\nu)& = \frac{4}{3\pi
 \Whittaker{\frac{3}{2}}{\frac{5}{2}}{\nu}} \left\{
 \frac{25}{2}\Whittaker{\frac{7}{2}}{\frac{5}{2}}{\nu}-
 \frac{1}{\nu^2}\Whittaker{\frac{5}{2}}{\frac{7}{2}}{\nu}+\frac{5}{2}\left(
 5-\frac{1}{\nu} \right)\Whittaker{\frac{7}{2}}{\frac{7}{2}}{\nu}
 \right.\nonumber\\
 &\left.+ \frac{21}{4}\Whittaker{\frac{9}{2}}{\frac{9}{2}}{\nu} -
 \frac{3}{2}\frac{\Whittaker{\frac{5}{2}}{\frac{5}{2}}{\nu}}{\Whittaker{\frac{3}{2}}{\frac{5}{2}}{\nu}}
 \left[
 5\Whittaker{\frac{5}{2}}{\frac{5}{2}}{\nu}+\left(5-\frac{1}{\nu}\right)
 \Whittaker{\frac{5}{2}}{\frac{7}{2}}{\nu}+\frac{3}{2}\Whittaker{\frac{7}{2}}{\frac{9}{2}}{\nu}\right]\right\},\\
 b'(\nu)&= -\frac{3}{2\Whittaker{\frac{3}{2}}{\frac{5}{2}}{\nu}}
 \left\{ \frac{\nu}{2}\Whittaker{\frac{1}{2}}{\frac{5}{2}}{\nu} +
 \frac{\nu}{3}\Whittaker{\frac{3}{2}}{\frac{5}{2}}{\nu} + \left[
 1+ \frac{3}{2}\nu
 \frac{\Whittaker{\frac{5}{2}}{\frac{5}{2}}{\nu}}{\Whittaker{\frac{3}{2}}{\frac{5}{2}}{\nu}}\right]
 \left[
 \Whittaker{-\frac{1}{2}}{\frac{5}{2}}{\nu}-\frac{2}{3}\Whittaker{\frac{1}{2}}{\frac{5}{2}}{\nu}\right]\right\},\\
 c'(\nu)&=
 b'(\nu)+\frac{4}{3}\frac{\Whittaker{\frac{1}{2}}{\frac{3}{2}}{\nu}}{\Whittaker{\frac{3}{2}}{\frac{5}{2}}{\nu}}+\frac{4}{3}\frac{\nu^2}{\left[\Whittaker{\frac{3}{2}}{\frac{5}{2}}{\nu}\right]^2}
 \left[ \frac{3}{2}\Whittaker{\frac{1}{2}}{\frac{3}{2}}{\nu}\Whittaker{\frac{5}{2}}{\frac{5}{2}}{\nu}-\frac{1}{2}\Whittaker{\frac{3}{2}}{\frac{3}{2}}{\nu}\Whittaker{\frac{3}{2}}{\frac{5}{2}}{\nu}\right].
\end{align}

\subsection{A channel ($^3P_1$)}

\begin{align}
 A(\nu)&= \frac{2}{3} A(\nu)|_{\text{Scalar}},\\
 a(\nu)&= \tilde{a}(\nu) +\frac{1}{\pi}\left[
 \frac{4}{3}+S_A(\nu) \right],\\
 S_A(\nu)&=
 -\frac{4}{3\Whittaker{\frac{3}{2}}{\frac{5}{2}}{\nu}}\left[
 \frac{3}{2} \Whittaker{\frac{5}{2}}{\frac{5}{2}}{\nu} + 2
 \Whittaker{\frac{7}{2}}{\frac{9}{2}}{\nu} + \left(
 1-\frac{3}{2\nu} \right)
 \Whittaker{\frac{5}{2}}{\frac{7}{2}}{\nu} \right],\\
 b(\nu)&= -\frac{3}{2}\nu
 \frac{\Whittaker{-\frac{1}{2}}{\frac{5}{2}}{\nu}}{\Whittaker{\frac{3}{2}}{\frac{5}{2}}{\nu}},\\
 c(\nu)&=
 b(\nu)+\frac{4\nu}{3}\frac{\Whittaker{\frac{1}{2}}{\frac{3}{2}}{\nu}}{\Whittaker{\frac{3}{2}}{\frac{5}{2}}{\nu}},\\
 A'(\nu)&= \frac{2}{3}A'(\nu)|_{\text{Scalar}},\\
 a'(\nu)&= \tilde{a}'(\nu)+\frac{1}{\pi}S_A'(\nu),\\
 S_A'(\nu)&= \frac{4}{3\Whittaker{\frac{3}{2}}{\frac{5}{2}}{\nu}}
 \left\{
 \frac{15}{4}\Whittaker{\frac{7}{2}}{\frac{5}{2}}{\nu}-\frac{3}{2\nu^2}\Whittaker{\frac{5}{2}}{\frac{7}{2}}{\nu}
 + \frac{5}{2}\left(1-\frac{3}{2\nu}\right)
 \Whittaker{\frac{7}{2}}{\frac{7}{2}}{\nu} + 7
 \Whittaker{\frac{9}{2}}{\frac{9}{2}}{\nu} \right. \nonumber\\
 &\left.
 -\frac{3}{2}
 \frac{\Whittaker{\frac{5}{2}}{\frac{5}{2}}{\nu}}{\Whittaker{\frac{3}{2}}{\frac{5}{2}}{\nu}}
 \left[ \frac{3}{2}\Whittaker{\frac{5}{2}}{\frac{5}{2}}{\nu} + \left(
 1-\frac{3}{2\nu} \right)
 \Whittaker{\frac{5}{2}}{\frac{7}{2}}{\nu} + 2
 \Whittaker{\frac{7}{2}}{\frac{9}{2}}{\nu} \right] \right\},\\
 b'(\nu)&=
 \frac{3}{2\Whittaker{\frac{3}{2}}{\frac{5}{2}}{\nu}}\left\{
 \frac{\nu}{2}\Whittaker{\frac{1}{2}}{\frac{5}{2}}{\nu} + \left[
 1+\frac{3}{2}\nu
 \frac{\Whittaker{\frac{5}{2}}{\frac{5}{2}}{\nu}}{\Whittaker{\frac{3}{2}}{\frac{5}{2}}{\nu}}\right]
 \Whittaker{-\frac{1}{2}}{\frac{5}{2}}{\nu}\right\},\\
 c'(\nu)&= b'(\nu)+\frac{4}{3}\frac{\Whittaker{\frac{1}{2}}{\frac{3}{2}}{\nu}}{\Whittaker{\frac{3}{2}}{\frac{5}{2}}{\nu}}+\frac{4}{3}\frac{\nu^2}{\left[\Whittaker{\frac{3}{2}}{\frac{5}{2}}{\nu}\right]^2}
 \left[ \frac{3}{2}\Whittaker{\frac{1}{2}}{\frac{3}{2}}{\nu}\Whittaker{\frac{5}{2}}{\frac{5}{2}}{\nu}-\frac{1}{2}\Whittaker{\frac{3}{2}}{\frac{3}{2}}{\nu}\Whittaker{\frac{3}{2}}{\frac{5}{2}}{\nu}\right].
\end{align}

\section{Spectral function for the continuum part}
\label{app:cont}
We summarize the continuum part of the phenomenological spectral function
$\text{Im}\Pi^{J,\text{pert}}(s)$ which are taken from the perturbative QCD
calculation up to $\mathcal{O}(\alpha_s)$ shown in Ref.~\cite{reinders85}.
Here,
\begin{equation}
 u = \sqrt{1-\frac{4m_h^2}{s}},
\end{equation}
and
\begin{equation}
 \Delta = \frac{2\alpha_s}{\pi}\ln 2.
\end{equation}
Then,
\begin{align}
 \text{Im}\tilde{\Pi}^{V,\text{pert}}(s)&= \frac{u(3-u^2)}{8\pi}
 \left[
 1+\frac{4}{3}\alpha_s \left\{
 \frac{\pi}{2u}-\frac{u+3}{4}
 \left(\frac{\pi}{2}-\frac{3}{4\pi}\right)
 \right\}
 \right]
 -\frac{3(1-u^2)^2}{8\pi u}\Delta\\
 \text{Im}\tilde{\Pi}^{P,\text{pert}}(s)&=
 \frac{3u}{8\pi}
 \left[
 1+ \frac{4\alpha_s}{3\pi u}
 \left\{
 \pi u
 \left[
 \frac{\pi}{2u}-\frac{3+u}{4}
 \left(
 \frac{\pi}{2}-\frac{3}{4\pi}
 \right)
 \right]
 +u
 -\frac{3(u^6-7u^4+19u^2+3)}{16(3-u^2)}\ln\frac{1+u}{1-u}
 \right.\right. \nonumber\\
 &\left.\left.
 +\frac{3u(11-4u^2+u^4)}{8(3-u^2)}
 \right\}
 \right] - \frac{3}{8\pi}\Delta \frac{1-u^2}{u}\label{eq:p}\\
 \text{Im}\tilde{\Pi}^{S,\text{pert}}(s)&=\frac{3u^3}{8\pi}
 \left[
 1+\frac{4\alpha_s}{3\pi u^3}
 \left\{
 \pi u^3
 \left[
 \frac{\pi}{2u}-\frac{1+u}{2}
 \left(
 \frac{\pi}{2}-\frac{3}{\pi}
 \right)
 \right]
 +
 \left(\frac{15}{16}-\frac{3}{8}u^2-\frac{33}{16}u^4\right)\ln\frac{1+u}{1-u}
 \right.\right. \nonumber\\
 & \left.\left. -\frac{15}{8}u+\frac{33}{8}u^3
 \right\}
 \right]
 - \frac{9u(1-u^2)}{8\pi}\Delta\\
 \text{Im}\tilde{\Pi}^{A,\text{pert}}(u)&=
 \frac{u^3}{4\pi}
 \left[
 1+\frac{4\alpha_s}{3\pi u^3}
 \left\{
 \pi u^3
 \left[
 \frac{\pi}{2u}-\frac{1+u}{2}
 \left(
 \frac{\pi}{2}-\frac{3}{\pi}
 \right)
 \right]
 +u^3+\frac{3(15-7u^2-7u^4-u^6)}{32}\ln\frac{1+u}{1-u}
 \right.\right. \nonumber\\
 & \left.\left. + \frac{3(u^5-2u^3-15)}{16} \right\}\right]
 - \frac{3u(1-u^2)}{4\pi}\Delta
 \end{align}
\end{widetext}

\section{List of obtained spectral parameters}
\label{app:param_list}

We summarize the results of the spectral parameters of the model
spectral function at various temperatures for which we compared them with the
lattice QCD results of the imaginary time correlators in Sec.~\ref{sec:IMC}.
In the case of $\Gamma=0$, the minimum $\chi^2$ will provide a good
estimation of the best Borel curve since $M^2_{\text{min}}$ is fixed.
When $\Gamma > 0$, however, this will no longer hold to choose the best
one among curves with different $\sqrt{s_0}$. Therefore, we used this
criterion only to choose the best $\Gamma$ value among a fixed $\sqrt{s_0}$
case for each temperature. In Tables
\ref{tbl:sumruleresult_jpsi}-\ref{tbl:sumruleresult_chic1}, we list sets
of resultant parameters obtained in this way (not all cases: we computed
more cases of $\sqrt{s_0}$ but we include the case of the smallest
$\sqrt{s_0}$ below which no Borel window is available).
We list $\sqrt{\chi^2}$ instead of $\chi^2$ since it is related to a crude
estimate of a systematic error in the mass as discussed in
Sec.~\ref{subsec:method}.

\begin{table*}[t]
 \caption{Spectral parameters for $J/\psi$ at finite
 temperature obtained from QCD sum rules.}
 \label{tbl:sumruleresult_jpsi}
 \begin{ruledtabular}
  \begin{tabular}{ccccccc}
   $T/T_c$ &$\sqrt{s_0}$ [GeV]&$m$ [GeV]&$\Gamma$ [MeV]&$f$ or
   $f_0$ [GeV$^2$]&$M_0^2$
   [GeV$^2$]& $\sqrt{\chi^2}$ [MeV] \\\hline
   1.00&3.26&2.975&0  &0.264&2.350&51.3\\
       &3.4 &3.010&0  &0.326&1.841&31.9\\
       &3.5 &3.026&0  &0.360&1.673&32.1\\
       &3.6 &3.055&32 &0.130&1.666&43.8\\\hline
   1.04&3.16&2.920&0  &0.211&1.751&18.3\\
       &3.28&2.947&0  &0.258&1.375&11.6\\
       &3.4 &2.997&58 &0.105&1.452&19.7\\
       &3.5 &3.035&100&0.123&1.487&29.2 \\\hline
   1.07&3.05&2.868&1  &0.0494&1.009&0.04 \\
       &3.2 &2.935&64 &0.0770&1.322&2.4 \\
       &3.4 &3.024&158&0.118 &1.509&9.4 \\
       &3.6 &3.100&240&0.158 &1.554&24.6 \\\hline
   1.09&3.04&2.901&126&0.0586&1.024&0.48 \\
       &3.2 &2.950&134&0.0838&1.254&1.2 \\
       &3.4 &3.037&216&0.125 &1.560&2.9 \\
       &3.6 &3.121&306&0.169 &1.626&13.1 \\
  \end{tabular}
 \end{ruledtabular}
\end{table*}

\begin{table*}[ht]
 \caption{Spectral parameters for $\eta_c$ obtained from QCD sum
 rules.}
 \label{tbl:sumruleresult_etac}
 \begin{ruledtabular}
  \begin{tabular}{ccccccc}
   $T/T_c$ &$\sqrt{s_0}$ [GeV]&$m$ [GeV]&$\Gamma$ [MeV]&$f$ or
   $f_0$ [GeV$^2$]&$M_0^2$
   [GeV$^2$]& $\sqrt{\chi^2}$ [MeV] \\\hline
   1.00&3.2 &2.915&0 &0.262&1.732&43.3 \\
       &3.3 &2.936&0 &0.307&1.413&31.1 \\
       &3.4 &2.950&0 &0.340&1.287&33.6 \\
       &3.5 &2.975&26&0.122&1.281&48.2 \\
       &3.6 &2.992&42&0.133&1.266&67.5 \\ \hline
   1.04&3.02&2.834&18 &0.0527&1.008&0.15 \\
       &3.2 &2.920&104&0.0918&1.243&2.1 \\
       &3.4 &3.011&202&0.140 &1.376&7.6 \\
       &3.6 &3.089&288&0.187 &1.427&21.0 \\\hline
   1.07&3.02&2.880&230&0.0713&1.022&0.78 \\
       &3.1 &2.900&198&0.0827&1.186&2.6 \\
       &3.2 &2.942&216&0.104 &1.367&3.8 \\
       &3.3 &2.987&252&0.127 &1.548&4.5 \\
       &3.4 &3.033&296&0.152 &1.794&4.9 \\
       &3.5 &3.079&342&0.179 &2.133&5.4 \\ \hline
   1.09&3.02&2.917&428&0.0902&1.029&1.5 \\
       &3.1 &2.915&296&0.0920&1.207&4.7 \\
       &3.2 &2.952&284&0.111 &1.374&7.1 \\
       &3.3 &2.995&306&0.134 &1.540&8.8 \\
       &3.4 &3.041&344&0.159 &1.710&10.2 \\
       &3.5 &3.088&390&0.186 &1.899&11.2 \\
  \end{tabular}
 \end{ruledtabular}
\end{table*}

\begin{table*}[ht]
 \caption{Summary of spectral parameters for $\chi_{c0}$ obtained from QCD sum
 rules.}
 \label{tbl:sumruleresult_chic0}
 \begin{ruledtabular}
  \begin{tabular}{ccccccc}
   $T/T_c$ &$\sqrt{s_0}$ [GeV]&$m$ [GeV]&$\Gamma$ [MeV]&$f$ or
   $f_0$ [GeV$^2$]&$M_0^2$
   [GeV$^2$]& $\sqrt{\chi^2}$ {MeV} \\\hline
   1.00&3.6 &3.288&0  &0.205 &2.230&16.5 \\
       &3.7 &3.314&0  &0.240 &2.094&15.1 \\
       &3.8 &3.368&40 &0.0924&2.170&22.6 \\ \hline
   1.04&3.28&3.083&0  &0.0901&1.367&0.31 \\
       &3.4 &3.156&46 &0.0420&1.537&2.8 \\
       &3.5 &3.215&80 &0.0548&1.651&6.6 \\
       &3.6 &3.270&110&0.0687&1.735&12.0 \\ \hline
   1.07&3.34&3.166&194&0.0428&1.428&0.15 \\
       &3.4 &3.199&200&0.0496&1.599&0.67 \\
       &3.5 &3.254&214&0.0630&1.691&2.5 \\
       &3.6 &3.297&218&0.0768&1.592&8.1 \\ \hline
   1.09&3.26&3.167&340&0.0404&1.331&0.05 \\
       &3.4 &3.226&296&0.0549&1.548&0.76 \\
       &3.5 &3.279&296&0.0672&1.891&0.47 \\
       &3.6 &3.329&300&0.0833&1.724&2.7 \\
  \end{tabular}
 \end{ruledtabular}
\end{table*}

\begin{table*}[ht]
 \caption{Summary of spectral parameters for $\chi_{c1}$ obtained from QCD sum
 rules.}
 \label{tbl:sumruleresult_chic1}
 \begin{ruledtabular}
  \begin{tabular}{ccccccc}
   $T/T_c$ &$\sqrt{s_0}$ [GeV]&$m$ [GeV]&$\Gamma$ [MeV]&$f$ or
   $f_0$ [GeV$^2$]&$M_0^2$
   [GeV$^2$]& $\sqrt{\chi^2}$ [MeV] \\\hline
   1.00&3.5 &3.349&0  &0.128 &2.382&22.2 \\
       &3.6 &3.364&0  &0.144 &2.355&17.3 \\
       &3.7 &3.386&6  &0.0520&2.331&20.7 \\
       &3.8 &3.434&46 &0.0615&2.428&28.9 \\ \hline
   1.04&3.2 &3.170&4  &0.0215&1.486&0.14 \\
       &3.3 &3.214&44 &0.0267&1.621&2.6 \\
       &3.4 &3.257&76 &0.0327&1.747&7.0 \\
       &3.5 &3.301&104&0.0397&1.866&12.6 \\
       &3.6 &3.344&130&0.0472&1.953&19.6 \\\hline
   1.07&3.2 &3.261&250&0.0295&1.489&0.03 \\
       &3.3 &3.285&252&0.0340&1.647&1.4 \\
       &3.5 &3.352&264&0.0466&1.904&8.4 \\
       &3.7 &3.432&286&0.0632&2.083&20.0 \\ \hline
   1.09&3.2 &3.328&422&0.0361&1.495&0.004 \\
       &3.3 &3.331&388&0.0392&1.664&0.66 \\
       &3.5 &3.384&362&0.0512&1.946&5.7 \\
       &3.7 &3.461&366&0.0681&2.140&15.1 \\
  \end{tabular}
 \end{ruledtabular}
\end{table*}

\begin{table*}[ht]
 \caption{Summary of spectral parameters for $\Upsilon$ obtained from QCD sum
 rules.}
 \label{tbl:sumruleresult_upsilon}
 \begin{ruledtabular}
  \begin{tabular}{ccccccc}
   $T/T_c$ &$\sqrt{s_0}$ [GeV]&$m$ [GeV]&$\Gamma$ [MeV]&$f$ or
   $f_0$ [GeV$^2$]&$M_0^2$
   [GeV$^2$]& $\sqrt{\chi^2}$ [MeV]\\\hline
   1.15&10.04&9.369&0  &1.670 &16.52&44.3 \\
       &10.16&9.401&0  &1.845 &14.86&31.7 \\
       &10.28&9.428&0  &1.959 &13.71&25.0 \\
       &10.34&9.440&0  &2.023 &13.26&24.9 \\
       &10.44&9.480&40 &0.694 &13.31&29.3 \\ \hline
   1.54&10.0 &9.350&0  &1.613 &16.20&38.3 \\
       &10.1 &9.377&0  &1.739 &14.74&27.9 \\
       &10.2 &9.401&0  &1.857 &13.68&21.6 \\
       &10.34&9.441&22 &0.647 &13.03&23.0 \\
       &10.5 &9.516&108&0.738 &13.46&30.3 \\
  \end{tabular}
 \end{ruledtabular}

 \caption{Summary of spectral parameters for $\eta_b$ obtained from QCD sum
 rules.}
 \label{tbl:sumruleresult_etab}
 \begin{ruledtabular}
  \begin{tabular}{ccccccc}
   $T/T_c$ &$\sqrt{s_0}$ [GeV]&$m$ [GeV]&$\Gamma$ [MeV]&$f$ or
   $f_0$ [GeV$^2$]&$M_0^2$
   [GeV$^2$]& $\sqrt{\chi^2}$ [MeV]\\\hline
   1.15&9.98 &9.315&0  &1.769 &15.71&48.6 \\
       &10.1 &9.347&0  &1.941 &13.48&35.1 \\
       &10.22&9.373&0  &2.094 &12.33&28.2 \\
       &10.28&9.385&0  &2.163 &11.91&28.3 \\
       &10.4 &9.430&44 &0.753 &11.92&34.7 \\ \hline
   1.54&9.94 &9.295&0  &1.705 &15.18&39.5 \\
       &10.04&9.322&0  &1.848 &13.23&28.9 \\
       &10.14&9.345&0  &1.977 &12.17&22.9 \\
       &10.28&9.387&28 &0.694 &11.69&25.7 \\
       &10.4 &9.441&86 &0.768 &11.95&31.8 \\
  \end{tabular}
 \end{ruledtabular}

 \caption{Summary of spectral parameters for $\chi_{b0}$ obtained from QCD sum
 rules.}
 \label{tbl:sumruleresult_chib0}
 \begin{ruledtabular}
  \begin{tabular}{ccccccc}
   $T/T_c$ &$\sqrt{s_0}$ [GeV]&$m$ [GeV]&$\Gamma$ [MeV]&$f$ or
   $f_0$ [GeV$^2$]&$M_0^2$
   [GeV$^2$]& $\sqrt{\chi^2}$ [MeV]\\\hline
   1.15&10.43&9.838&0  &0.609 &13.49&22.5 \\
       &10.58&9.882&0  &0.696 &12.81&13.4 \\
       &10.73&9.943&22 &0.254 &12.62&16.5 \\
       &10.8 &9.984&46 &0.274 &12.83&19.9 \\ \hline
   1.54&10.34&9.773&0  &0.542 &12.36&9.13 \\
       &10.52&9.839&16 &0.207 &11.89&7.17 \\
       &10.72&9.960&84 &0.260 &12.58&14.5 \\
       &10.8 &10.007&110&0.283 &12.83&18.1 \\
  \end{tabular}
 \end{ruledtabular}

 \caption{Summary of spectral parameters for $\chi_{b1}$ obtained from QCD sum
 rules.}
 \label{tbl:sumruleresult_chib1}
 \begin{ruledtabular}
  \begin{tabular}{ccccccc}
   $T/T_c$ &$\sqrt{s_0}$ [GeV]&$m$ [GeV]&$\Gamma$ [MeV]&$f$ or
   $f_0$ [GeV$^2$]&$M_0^2$
   [GeV$^2$]& $\sqrt{\chi^2}$ [MeV]\\\hline
   1.15&10.18&10.043&0  &0.445 &13.53&1.9 \\
       &10.26&10.049&0  &0.460 &13.52&2.0 \\
       &10.34&10.087&34 &0.157 &13.88&3.9 \\
       &10.4 &10.114&56 &0.165 &14.13&5.6 \\ \hline
   1.54&10.1 &10.005&20 &0.133 &12.76&0.02 \\
       &10.22&10.063&72 &0.148 &13.36&1.5 \\
       &10.34&10.114&114&0.163 &13.87&3.7 \\
       &10.4 &10.142&136&0.172 &14.13&5.4 \\
  \end{tabular}
 \end{ruledtabular}
\end{table*}


\end{document}